\documentclass[preprint,12pt,epsf]{aastex}
\usepackage{natbib}

\newcommand  \kms   {km~s$^{-1}$} 

\newcommand  \sii   {[\ion{S}{2}]}
\newcommand  \oiii  {[\ion{O}{3}]}
\newcommand  \ha    {H$\alpha$} 

\newcommand  \hii   {\ion{H}{2}}

\newcommand  \um    {$\mu$m}

\shorttitle{Young Massive Stars in N\,44}
\shortauthors{Chen et al.}

\begin{document}

\title{{\it Spitzer} View of Young Massive Stars in the LMC H\,II Complex
 N\,44}

\author{C.-H. Rosie Chen\altaffilmark{1,2}, You-Hua Chu\altaffilmark{2}
 and Robert A. Gruendl\altaffilmark{2}}
\affil{Department of Astronomy, University of Illinois,
 Urbana, IL 61801}
\email{c-chen@astro.uiuc.edu,chu@astro.uiuc.edu,gruendl@astro.uiuc.edu}

\author{Karl D. Gordon}
\affil{Space Telescope Science Institute, Baltimore, MD 21218}
\email{kgordon@stsci.edu}

\and

\author{Fabian Heitsch}
\affil{Department of Astronomy, University of Michigan, Ann Arbor, MI 48109}
\email{fheitsch@umich.edu}

\altaffiltext{1}{present address: Department of Astronomy, University
 of Virginia, Charlottesville, VA 22904; rchen@virginia.edu.}
\altaffiltext{2}{Visiting Astronomer, Cerro Tololo Inter-American Observatory.
 CTIO is operated by AURA, Inc.\ under contract to the National Science
 Foundation.}

\begin{abstract}

The \hii\ complex N\,44 in the Large Magellanic Cloud (LMC) provides an
excellent site to perform a detailed study of star formation in a mild 
starburst, as it hosts three regions of star formation at different
evolutionary stages and it is not as complicated and confusing as
the 30 Doradus giant \hii\ region.  We have obtained {\it Spitzer Space
Telescope} observations and complementary ground-based 4m $uBVIJK$
observations of N\,44 to identify candidate massive young stellar objects
(YSOs).  We further classify the YSOs into Types I, II, and III, according 
to their spectral energy distributions (SEDs).
In our sample of 60 YSO candidates, $\sim 65$\% of them are resolved
into multiple components or extended sources in high-resolution 
ground-based images.
We have modeled the SEDs of 36 YSOs that appear single or dominant 
within a group.  We find good fits for Types I and I/II YSOs,
but Types II and II/III YSOs show deviations between their observed
SEDs and models that do not include PAH emission.  We have also 
found that some Type III YSOs have central holes  in their disk 
components.
YSO counterparts are found in four ultracompact \hii\ regions and
their stellar masses determined from SED model fits agree well
with those estimated from the ionization requirements of the \hii\
regions.
The distribution of YSOs is compared with those of the underlying 
stellar population and interstellar gas conditions to illustrate
a correlation between the current formation of O-type stars and
previous formation of massive stars.  Evidence of triggered star
formation is also presented.

\end{abstract}


\section{Introduction}

Star formation frequently takes place in high concentrations of 
massive stars at intense levels, a phenomenon referred to as 
``starbursts.''
The massive stars in a starburst can inject energies into the
interstellar medium (ISM) to photoionize the ambient medium and
to dynamically sweep up the medium into expanding shells.
The expanding shells can trigger subsequent star formation either by
compressing ambient dense clouds or by collect-and-collapse within the
shell \citep{El98}.  Starbursts can easily spread over areas
10$^2$-10$^3$ pc across, and play a vital role in determining the
large-scale structures of their host galaxies.

While starbursts are the most prominent features in a galaxy, their 
detailed properties cannot be easily studied: in distant galaxies the 
stellar content is not resolved and in the Milky Way the distances and 
association among stars in a starburst are uncertain.  The Large and 
Small Magellanic Clouds (LMC \& SMC; MCs) are the only galaxies in
which stars are at common, known distances and can be individually 
resolved down to $\sim$1 $M_\odot$ at optical wavelengths 
\citep{GBH06,Noetal06}.  Recent {\it Spitzer Space Telescope} observations 
in the mid-infrared (mid-IR) enabled the detection of individual young 
stellar objects (YSOs) in the MCs, revealing the on-going star formation 
\citep{CYetal05,JTetal05,CGC08}.  It is now possible to use the resolved 
stellar population to map out star formation as a function of space and 
time in the MCs.  As the ISM of the MCs has been surveyed in great detail, 
it is further possible to determine the relationship between star formation 
and the physical properties of the ISM.  Consequently, studies of starbursts 
in the MCs allow us to investigate fundamental issues about star formation, 
such as whether and how the stellar energy feedback from massive stars 
triggers the formation of next-generation stars, whether the initial 
mass function (IMF) depends on the interstellar conditions for star 
formation, and whether massive stars are formed only under specific 
conditions \citep{ZY07}.

We have chosen the \hii\ complex LH$\alpha$ 120-N\,44 \citep[N\,44;][]{He56}
in the LMC to carry out a comprehensive study of star formation in a mild
starburst situation, as it is one of the top ranking \hii\ complexes in
the LMC but is not as complicated and confusing as the giant \hii\ region
30 Doradus.  N\,44 contains three OB associations, LH47, 48, and 49 
\citep{LH70}, that are in different evolutionary stages and interstellar
structures: LH47 in the central superbubble, LH48 in one contiguous \hii\ 
region at the northeast rim of the superbubble, and LH49 in a group of 
\hii\ regions to the southeast exterior of the superbubble 
(Figures.~\ref{fig:n44opt}a,b).
Along the western rim of the superbubble exist a number of dense \hii\ 
regions where star formation may have been triggered by the expansion 
of the superbubble \citep{OM95}.  Surveys of CO in the N\,44 complex
\citep[Fig.~\ref{fig:n44opt}d;][]{FYetal01} show a high concentration
of molecular gas at the western rim of the central superbubble, another
high concentration at the \hii\ regions around LH49, and a weaker 
concentration to the north of the superbubble where a couple of small \hii\ 
regions are visible but no OB associations are identified.

The effects of stellar energy feedback are evident in N\,44.  The 
\ha\ image (Fig.~\ref{fig:n44opt}a) shows the reach of stellar UV 
radiation.  The X-ray image (Fig.~\ref{fig:n44opt}c) reveals 10$^6$ K 
gas heated by fast stellar winds and supernova explosions.  While the 
expansion of the central superbubble may have triggered the intense star 
formation along its western rim, the hot gas outflow from the superbubble 
may be responsible for the onset of star formation at the northeastern 
extension of the south molecular peak.

To study the current star formation in N\,44, we have obtained 
{\it Spitzer} mid-IR and complementary ground-based optical and
near-IR observations.
These observations have been analyzed and the results are reported
in this paper.
Section 2 describes the observations and data reduction.
Section 3 reports the initial identification of YSO candidates and 
Section 4 describes a further pruning of the YSO candidate list.
In Section 5 we classify the YSOs, and in Section 6 we determine their 
physical properties by modeling their spectral shapes.
In Section 7 we discuss the massive star formation properties.
A summary is given in Section~8.

\section{Observations and Data Reduction}

We have obtained {\it Spitzer} mid-IR observations to diagnose YSOs.
To extend the spectral energy distribution (SED) and to improve the angular 
resolution, we have further obtained ground-based optical and near-IR 
imaging observations of N\,44.  
We have also retrieved available images in the {\it Hubble Space Telescope} 
({\it HST}) archive to examine the optical counterparts and environments 
of the YSOs.

\subsection{{\it Spitzer} IRAC and MIPS Observations}

Our {\it Spitzer} observations of N\,44 were made with the 
InfraRed Array Camera \citep[IRAC;][]{FGetal04} on 2005 March 27
and the Multiband Imaging Photometer for {\it Spitzer} 
\citep[MIPS;][]{RGetal04} on 2005 April 7.
The IRAC observations were obtained in the 3.6, 4.5, 5.8, and 8.0 
\um\ bands using the mapping mode to cover a $\sim30' \times 30'$ area
in each band.  The pixel size was 1\farcs2~pixel$^{-1}$, and the common
area covered by all four bands is $\sim30' \times 23'$.  
At each pointing, exposures were made with a five-point cyclic dithering 
pattern in the 30~s high dynamic range mode.
The total integration time at each pointing is $\sim$ 150 s.
The mosaicked maps were processed with the Spitzer Science Center (SSC) 
pipeline (ver.\ S11.4.0) and provided to us as part of the Post Basic 
Calibrated Data (PBCD) products.

The MIPS observations were made in the 24, 70, and 160 \um\ bands 
using the scan map mode at the medium scan rate.
The mapping consists of sixteen $0\fdg5$ scan legs with a 
cross-scan step of 148$''$ to cover a region of 
$20' \times 30'$ in all three MIPS bands.
The MIPS DAT version 3.00 \citep{GKetal05} was used for the
basic processing and final mosaicking of the individual images.
In addition, the 24 \um\ image has been corrected for a readout
offset and divided by a scan-mirror-independent flat field, and
the 70 and 160 \um\ images have been corrected for a pixel-dependent
background using a low-order polynomial fit to the source-free regions.
The final mosaicked maps have pixel scales of 1\farcs25, 9\farcs85, and 
16\farcs0~pixel$^{-1}$, and exposure times of roughly 160, 80, and 16 s 
at 24, 70, and 160 \um, respectively.

The common area covered by all IRAC and MIPS bands is $20' \times 23'$.
To achieve a more complete view of the  molecular cloud to the north of
N\,44 superbubble, we have used the {\it Spitzer} survey of the LMC 
\citep[SAGE,][]{MMetal06} to extend our IRAC 3.6 and 5.8 \um\ data in 
the declination direction.
The final IRAC and MIPS field we have analyzed is $20' \times 25'$.

To carry out aperture photometry for point sources in the IRAC 
images we use the IRAF package \texttt{apphot}.
First, the sources were identified with the automated source finding 
routine \texttt{daofind}, using parameters optimized to find the majority 
of point-like sources while minimizing the inclusion of peaks of extended
dust emission.
In regions containing multiple sources superposed on extended emission,
\texttt{daofind} does not always identify the same sources at the
same locations in the four IRAC bands.
Therefore, we identified sources in each of the four IRAC bands and
merged the four source lists into a master source list using 1\farcs2 
(= 1 pixel) as the criterion for coincidence. 
The resulting master list is then used for photometric measurements in 
all four bands.
The photometric measurements were made with a source aperture of 
3\farcs6 (3-pixels) radius and an annular background aperture at
radii of 3\farcs6--8\farcs4 (3--7 pixels).
Finally, an aperture correction was applied, and the fluxes were 
converted into magnitudes using the correction factors and zero-magnitude
fluxes provided in the IRAC Data Handbook and listed in Table~\ref{photpar}.

The photometric measurements from long- and short-exposure IRAC
observations were averaged with weights proportional to the inverse
square of their errors. 
For sources that were saturated in the long exposures, their 
measurements from the short exposures were adopted.
The results were used to produce a photometric catalog of 17,002
IRAC sources in the $20' \times 25'$ field of N\,44. 
This catalog is presented as an ASCII table and an example is 
shown in Table~\ref{photcat}.

MIPS images have lower angular resolution.  Source identification
using \texttt{daofind} is feasible only for the 24 \um\ images.
In the 70 and 160 \um\ images, point sources cannot be easily 
resolved from one another or from a bright diffuse background;
therefore, the few apparent point sources were identified by
visual inspection.
The photometric measurements were made with parameters appropriate
for the point spread functions (PSFs), as recommended in the MIPS 
Data Handbook and given in Table~\ref{photpar}.
Note that the aperture corrections adopted for 70 and 160 \um\ measurements 
are those for sources of temperatures 15 and 10 K, respectively.
For sources of higher temperatures, such as 1000--3000 K, the adopted
aperture corrections will result in fluxes 8\% too high in 70 \um\ 
and 2\% too high in 160 \um\ because the source emission peaks at 
shorter wavelengths; however, these errors do not significantly affect 
the analysis and conclusions of our study of YSOs.
In the 160 \um\ image, only one point source can be clearly identified;
therefore, the 160 \um\ photometry will be discussed in the text but
not included in the photometric catalog.
The catalogs of the MIPS 24 and 70 \um\ bands are merged with the 
IRAC catalog and included in Table~\ref{photcat}.

The {\it Spitzer} images of N\,44 in the 3.6, 8.0, and 24 \um\ bands
are shown in Figure~\ref{fig:n44img}.
The 3.6 \um\ image is dominated by stellar emission, the 8.0
\um\ image shows the polycyclic aromatic hydrocarbon (PAH) emission,
while the 24 \um\ image is dominated by dust continuum emission
\citep{LD01,LD02,DL07}.
To better illustrate the relative distribution of emission in the 
different bands, we have produced a color composite with 3.6, 8.0, 
and 24 \um\ images mapped in blue, green, and red, respectively.
In this color composite, shown in Fig.~\ref{fig:n44img}d, dust emission 
appears red and diffuse, stars appear as blue point sources, red supergiants
appear yellow, and dust-shrouded YSOs and AGB stars appear red.

\subsection{CTIO 4~m ISPI and MOSAIC Observations}

We obtained near-IR images in the $J$ and $K_s$ bands with the Infrared Side 
Port Imager (ISPI) on the Blanco 4~m telescope at Cerro Tololo Inter-American 
Observatory (CTIO) on 2005 November 14--15.
The images were obtained with the $2{\rm K} \times 2$K HgCdTe HAWAII-2 
array, which had a pixel scale of 0\farcs3~pixel$^{-1}$ and a 
field-of-view of $10\farcm25 \times 10\farcm25$.
Six fields were observed to map N\,44.
Each field was observed with ten 30 s exposures in the $J$ band and 
twenty 30 s exposures in the $K_s$ band (each of the latter was
coadded from two 15~s frames to avoid background saturation).
The observations were dithered to aid in the removal
of transients and chip defects.
Owing to the diffuse emission in N\,44, we obtained sky frames at $\sim20'$ 
south of N\,44 to aid in sky subtractions.
The sky observations were made before and after each set of ten on-source
exposures.
All images were processed using the IRAF package \texttt{cirred} for 
dark and sky subtraction and flat-fielding.
The astrometry of individual processed images was solved with the 
routine \texttt{imwcs} in the package \texttt{wcstools}.
The astrometrically calibrated images are then coadded to produce a total 
exposure map for each filter.
The flux calibration was carried out using 2MASS photometry of 
isolated sources.

We obtained the SDSS $u$ and Johnson-Cousins $BVI$ broadband images of 
N\,44 with the MOSAIC II CCD Imager on the CTIO Blanco 4~m telescope on 
2006 February 2.
The MOSAIC Imager consists of eight $2{\rm K} \times 4$K SITe CCDs.
The CCDs have a pixel scale of 0\farcs27~pixel$^{-1}$, yielding a 
total field-of-view of $36'\times36'$.
The entire N\,44 was imaged in a single field.
All $uBVI$ images have been processed with the standard reduction procedure:
bias and dark were subtracted, flat-fielding was applied, and multiple
frames in each filter were combined to remove cosmic rays and improve the S/N.
The astrometry for the processed images were performed by referencing
to stars in the USNO B1.0 catalog.
The flux calibration was carried out using photometric measurements of 
isolated sources in \citet{Ch07} and \citet{ZDetal04}.

\subsection{Archival {\it HST} Images}

We have searched the {\it HST} archive for Wide Field Planetary Camera 2 
(WFPC2) images in the field of N\,44.
The available observations are listed in Table~\ref{wfpc2obs},
in which the coordinates, program ID, PI, filter, and exposure 
time are given.
Most of the observations contain multiple exposures for the same pointing 
and filter, and these images are combined using the IRAF routine 
\texttt{crrej} to remove cosmic rays and produce a total exposure map.
The astrometry was refined for each resultant image by referencing to stars 
in the USNO B1.0 catalog.

Seven fields have observations, but not all are useful; for example,
the wide $U$ band (F300W) images have very low S/N.
The most useful images are those taken with the \ha\ (F656N) 
or Str\"omgren $y$ (F547M) filter.
The former shows ionized gas and the latter shows stars at high resolution.

\subsection{Other Useful Archival Datasets}

To construct SEDs for the sources in our {\it Spitzer} photometric 
catalog of N\,44, we have expanded the catalog by adding $UBV$ photometric 
measurements obtained from CCD images taken with the CTIO 0.9~m telescope
\citep{Ch07}.
In regions not covered by these 0.9~m $UBV$ observations, we use 
the $UBVI$ photometry from the Magellanic Cloud Photometric 
Survey \citep[MCPS,][]{ZDetal04}.
We have also added near-IR data from the Point Source Catalog of the 
Two Micron All Sky Survey \citep[2MASS,][]{SMetal06}.

When merging the datasets, we allow a 1$''$ error margin for matching
{\it Spitzer} sources with optical or near-IR sources.
The final catalog lists each source's right ascension, declination,
and magnitudes in the order of increasing wavelengths, i.e.,
$U$, $B$, $V$, $I$, $J$, $H$, $K_s$, 3.6, 4.5, 5.8, 8.0, 24, and 70 \um.
The entire catalog is presented as an ASCII table and an example 
is shown in Table~\ref{photcat}.
These magnitudes can be converted to flux densities using the 
corresponding zero-magnitude flux listed in Tables~\ref{photpar} and 
\ref{flux2mag} and then used to construct SEDs for individual sources.

Finally, we have used \ha\ images of N\,44 from the
Magellanic Cloud Emission Line Survey \citep[MCELS,][]{SRetal99}
to examine the large-scale distribution of dense ionized gas
and to compare with images at other wavelengths.
As the angular resolution of this survey is $\sim2''$, we have
used additional \ha\ images taken with the CTIO 0.9 m telescope
by M.~A.~Guerrero \citep{NYetal02}.
These images, with a pixel size of 0\farcs4 pixel$^{-1}$, are used 
to show the immediate environments of YSOs.

\section{Initial Identification of Massive YSO Candidates}

\subsection{Expected Spectral Properties of YSOs}

YSOs have SEDs that differ from stars because they are shrouded in dust, 
which absorbs the stellar radiation and irradiates at IR wavelengths; 
therefore, it is possible to identify YSOs from their IR excesses.
It is, however, difficult to decipher the IR excess of massive YSOs
because the distribution of their circumstellar dust is not well known.
As their formation mechanism is still uncertain \citep[e.g.,][]{SPH00,ZY07},
it is not even known whether massive YSOs ubiquitously possess accretion 
disks \citep[e.g.,][]{CRetal07}.

We will use the commonly accepted physical structure and evolution
of low-mass YSOs as a starting point to decipher the SEDs of 
higher-mass YSOs.
Low-mass YSOs are believed to be initially surrounded by a small accretion 
disk and a large infalling envelope with bipolar cavities, and as they
evolve, the envelope and disk dissipate.
Different evolutionary stages result in different SEDs that have been 
used to classify low-mass YSOs, i.e., the Class I/II/III system \citep{La87}.
In this classification system, a Class I YSO has a compact accretion disk 
and a large infalling envelope with bipolar cavities; its SED  
is dominated by emission from the envelope and rises longward of 2 \um. 
A Class II YSO has dispersed most of its envelope and is surrounded by a 
flared disk; 
its SED, dominated by emissions from the central source and disk, 
is flat or falls longward of 2 \um. 
A Class III YSO has cleared most of the disk so its SED shows stellar
photospheric emission with little or no excess in near-IR.

These geometries of dust disk and envelope applicable to low-mass YSOs have
been adopted in radiative transfer models of high-mass YSOs
by a number of investigators \citep[e.g.,][]{WBetal04b,RTetal06}.
In their models, at an early evolutionary stage, a YSO has a small disk and
a large envelope with narrow bipolar conical cavities; its SED is dominated 
by the envelope emission and shows a generally rising trend from the 
shortest detectable wavelength to beyond 24 \um.
At an intermediate evolutionary stage, the opening angle of the bipolar 
cavity increases and the star and disk may be exposed; 
the YSO's SED thus shows double peaks: one below 1 \um\ (stellar 
emission) with its intensity dependent on the viewing angle, and 
the other rising longward of 1 \um\ but turning flat or falling 
shortward of 10--20 \um. 
At a late evolutionary stage when most of the envelope and disk have
been dispersed, the SED shows a bright stellar emission with a modest 
mid-IR excess.
These models provide useful links between circumstellar dust 
structures and SEDs; thus, we will use the general trends of 
the SEDs to identify YSO candidates in N\,44.

\subsection{Selection of Massive YSO Candidates}

Owing to their excess IR emission, YSOs are positioned in redder parts
of the color-color and color-magnitude diagrams (CMDs) than normal stars.
However, background galaxies and asymptotic giant branch (AGB) stars
can also be red sources, and these contaminants exist in non-negligible 
numbers.
To separate YSOs from these contaminants, we have examined several 
color-color diagrams and CMDs with solely IRAC bands as well as 
combinations of IRAC with $JHK_s$ or MIPS bands \citep{GC08}.
We find that for many sources the 2MASS catalog is too shallow to 
detect their counterparts in $JHK_s$, and the MIPS observations 
cannot resolve them from nearby sources or bright diffuse background.
To include most of the sources, we decide to concentrate on diagnostic 
diagrams using only the IRAC bands.
The IRAC [3.6]$-$[4.5] vs.\ [4.5]$-$[8.0] color-color diagram has been
suggested by \citet{Setal07} and the [8.0] vs.\ [4.5]$-$[8.0] CMD
has been suggested by \citet{HPetal06} to offer the best separation 
of YSOs from contaminants.
We have adopted the latter approach for the initial selection of
massive YSOs because galaxies and evolved stars have
different distributions in brightness and can be effectively
excluded using simple criteria.

Figure~\ref{fig:cmds} displays the [8.0] vs.\ ([4.5]$-$[8.0]) CMD 
of all sources detected in N\,44.
The sources in the prominent vertical branch centered at 
$([4.5]-[8.0]) \sim 0.0$ are mostly main-sequence, giant, and 
supergiant stars.
The contaminating background galaxies are concentrated in the
lower part of the CMD, and the AGB and evolved stars are
distributed mostly in the upper part of the CMD.
Below we discuss the criteria used to exclude these contaminating 
sources from the high-mass YSO candidates.

\subsubsection{Excluding Normal and AGB Stars}

Normal stars, i.e., main-sequence, giant, and supergiant stars, do 
not have excess IR emission and thus it is relatively easy to 
distinguish them from YSOs with a color criterion $([4.5]-[8.0]) < 0.5$ 
\citep{WBetal04a,HPetal06}.
On the other hand, evolved stars, e.g., AGB and post-AGB (pAGB) stars,
can have circumstellar dust and show excess IR emission and red colors.
To examine the locations of such evolved stars in the CMD, 
we first use known objects in the field of N\,44.
Using the SIMBAD database, we have found 10 confirmed AGB stars at 
various evolutionary stages, such as carbon stars, M-type variables, 
IR carbon stars, and OH/IR stars.
These 10 objects are marked by open squares in the [8.0] vs.\ [4.5]$-$[8.0] 
CMD (Fig.~\ref{fig:cmds}).
These sources have a color range of 0.2 $<$ ([4.5]$-$[8.0]) $<$ 1.3
and a magnitude range of 6.4 $<$ [8.0] $<$ 11.4.

We have also used models for Galactic C- and O-rich AGB stars \citep{Gr06}
to illustrate their expected locations in the CMD (Fig.~\ref{fig:cmds}).
To avoid crowding, we plot only models for a stellar luminosity of 3000 
$L_\odot$ to illustrate the range of colors.
For a luminosity range $1\times10^3$ to $6\times10^4 L_\odot$ \citep{Po93}, 
the expected loci of AGB stars in the CMD can move vertically by 1.2 to 
$-3.3$ mag.
As the chemistry of AGB atmospheres is dominated by nucleosynthesized material,
these Galactic models are good approximations for LMC objects although the
LMC metallicity is only $1/3$ solar.

We adopt the ([4.5]$-$[8.0]) $\ge 2.0$ criterion to exclude normal and 
AGB/pAGB stars.
As shown in Fig.~\ref{fig:cmds}, this criterion does exclude all known 
AGB/pAGB stars and a great majority of AGB models.
In our N\,44 field, 60 luminous sources are found with 0.5 $<$ 
([4.5]$-$[8.0]) $<$ 2.0 and [8.0] $\le$ 12.0, and indeed almost all
of them have SEDs consistent with those of AGB/pAGB stars, with the
remaining few appearing to be normal stars contaminated by
nebular background.

\subsubsection{Avoiding Background Galaxies \label{bggal}}

Background galaxies are concentrated in the lower part of the CMD 
in Fig.~3, bounded by ([4.5]$-$[8.0]) $> 0.5$ and 
[8.0] $\ge 14.0-$([4.5]$-$[8.0]), as suggested by \citet{HPetal06}
using the {\it Spitzer} Wide-Area Infrared Extragalactic Survey 
\citep[SWIRE,][]{LCetal03}.
We compare the surface density of sources bounded by 
[8.0] $\ge 14.0-$([4.5]$-$[8.0]) and [8.0] $\le$ 13.0
in the CMD of N\,44 to that of the SWIRE Survey.
For an area of $20'\times25'$, 284 sources are detected
in N\,44 within this wedge of CMD.
This surface density, 0.57 sources~arcmin$^{-2}$, is much
higher than that of SWIRE Survey, 0.06 sources~arcmin$^{-2}$,
although the SWIRE Survey had longer exposure times and thus
higher sensitivities \citep{GC08}.
This higher surface density of sources in N\,44 is most likely
attributed to a population of low-mass YSOs, as such YSOs are
expected to occupy this part of CMD for the LMC's distance
modulus of  $\sim$18.5 \citep{Fe99,WBetal04a,RTetal06}.
Therefore, as we adopt the criterion [8.0] $< 14.0-$([4.5]$-$[8.0]) 
to exclude background galaxies, we have also excluded YSOs with masses 
$\lesssim4$ $M_\odot$.

\subsubsection{Massive YSO Candidates}

After applying the two criteria $([4.5]-[8.0]) \ge 2.0$ and 
$[8.0] < 14 - ([4.5]-[8.0])$ to exclude most of normal and AGB stars
and background galaxies, we obtain a list of 99 YSO candidates.
To search for additional candidates that are more embedded and 
hence not detected in the 4.5 \um\ band, we resort to the 24 \um\
sources.  
Only one 24 \um\ source does not have a corresponding 4.5 \um\ source.
This source appears extended in the 8.0 \um\ image, indicating that
it is likely a small interstellar dust feature.
Therefore, no new objects are added to the list of 99 YSO candidates.

The 99 YSO candidates selected from the above CMD criteria still include 
a significant number of small dust features, obscured evolved stars, 
and bright background galaxies.
These contaminants need to be examined closely to assess their 
nature and need to be excluded from the YSO list.
In the next section we discuss how we use SEDs and multi-wavelength 
images to confirm and classify YSOs.

\section{Further Pruning of YSO Candidates}

To differentiate between YSOs and contaminants, we examine each YSO
candidate's morphology, environment, brightness, and SED to utilize 
as much information as possible in our consideration.
We have prepared the following multi-wavelength images with identical
field-of-view for each YSO candidate: MCELS \ha; MOSAIC $B$, $V$, and 
$I$; ISPI $J$ and $K_s$; IRAC 3.6, 4.5, 5.8, and 8.0 \um; and MIPS 
24 and 70 \um\ observations.
For some YSO candidates that have {\it HST} WFPC2 \ha\ and Str\"omgren
$y$ images available, these {\it HST} images replace the ground-based
\ha\ and $V$ images.
For YSO candidates that are not covered by our ISPI observations, 2MASS 
images are used.
In addition, we have constructed an SED for each source from optical
to 70 \um\ using the extended photometric catalog described in \S 2.4.
For each YSO candidate, we display and examine the multi-wavelength
images, its SED, and its location in the  [8.0] vs.\ ([4.5]$-$[8.0]) 
CMD simultaneously.
The nature of each YSO candidate is assessed by three of us (Chen, Chu,
and Gruendl) independently multiple times.
We gain experience from each round of examination and use our new
knowledge to aid in the next round of examination.
For most sources our classifications converge, but a few sources have
ambiguous properties and their classifications are thus uncertain.

\subsection{Identification of Background Galaxies}

Background galaxies can be identified from their morphologies if they
are resolved.
To diagnose unresolved background galaxies, we have to resort to
SEDs.
Galaxies co-located with YSO candidates in the [8.0] vs.\ ([4.5]$-$[8.0]) 
CMD are abundant in gas and dust, such as late-type galaxies or 
active galactic nuclei (AGN).
The SEDs of late-type galaxies are characterized by two broad humps, one
from stellar emission over optical and near-IR wavelength range and 
the other from dust emission over mid- to far-IR range.
The observed SED of an AGN depends on the viewing angle with respect to 
its dust torus; it can be flat from optical to far-IR or obscured in optical, 
showing only mid- to far-IR emission \citep{FAetal05,HEetal05,RRetal05}.  
In the latter case, the SED can be falling or rising at 24 \um.  
We further use the interstellar environment as a secondary diagnostic
for background galaxies, since galaxies are statistically less likely
to occupy preferred positions in prominent dust filaments.
Based on these considerations, two of our CMD-selected YSO candidates, 
sources 052042.0$-$674307.7 and 052106.8$-$675715.9, are reclassified as 
background galaxies.

\subsection{Identification of Evolved Stars}

The known AGB stars in N\,44 have SEDs peaking between 1 and 8 \um, 
as shown in Figure~\ref{fig:agbseds}. 
These SEDs are similar to those of Galactic AGB stars, corresponding to 
dust temperatures of $\sim 400-1,000$~K \citep{RRetal86}.
These known AGB stars in N\,44 are bluer than our selection criteria
for YSOs; however, the more obscured AGBs may have lower dust 
temperatures, exhibit redder colors, and occupy the same regions
as the YSO candidates in the CMD \citep[e.g.,][]{Betal06}.
We identify such obscured AGB or evolved stars based on their SEDs,
whose shapes are similar to those shown in Fig.~\ref{fig:agbseds} but
peaking at longer wavelengths.
We have also used the interstellar environment as a secondary criterion
to identify AGB and evolved stars, since evolved stars are not expected
to be located at preferred positions in diffuse dust emission.
From these considerations, two of our YSO candidates, 
sources 052221.0$-$680515.3 and 052351.1$-$675326.6, are reclassified as 
AGB/evolved stars.

\subsection{Identification of Dust Clumps}

Warm interstellar dust may show SEDs similar to those of 
circumstellar dust in YSOs.
As the angular resolution of IRAC images is $\sim 2''$, 
corresponding to 0.5 pc for a LMC distance of 50 kpc,
some small dust clumps may be identified as point sources
and included in the YSO candidate list.
Owing to our conservative compilation of master source list for
IRAC photometry (\S 2.1), a star projected near a dust clump
may also make its way into our YSO candidate list.
To identify these two types of YSO imposters, we use the ISPI
$JK_s$ images that have higher angular resolution.
In the ISPI $K_s$ images, stars or YSOs appear unresolved, while
dust clumps may appear as extended emission.
The $J - K_s$ color further differentiates between stars and YSOs,
as stars are brighter in $J$ and YSOs brighter in $K_s$.
Aided by the ISPI images, we find that 23 of our 99 YSO candidates 
are interstellar dust clumps, and another 12 are stars projected
near dust clumps.

\subsection{Final Massive YSO Sample}

The results of our examination of 99 YSO candidates are given in 
Table~\ref{ysoclass}, which lists source name, 
ranking of the brightness at 8 \um, magnitudes in the 
$U$, $B$, $V$, $I$, $J$, $H$, $K_s$, 3.6, 4.5, 5.8, 8.0, 24, 
and 70 \um\ bands, source classification, and remarks.
Note that some of the YSOs appear as single sources in IRAC 
images but are resolved into multiple sources in ISPI and MOSAIC 
images; for these sources, the photometric measurements 
in Table~\ref{ysoclass} were made for the dominant YSO sources
and these entries are remarked in Column ``Flag''.

After excluding background galaxies, evolved/AGB stars, and dust
clump imposters, 60 YSO candidates remain.  As these are most 
likely bona fide YSOs, we will simply call them YSOs in the rest
of the paper.

\subsection{Comparison with YSO Samples Identified by Others}

YSOs in N\,44 have been identified using other methodologies
by, e.g., \citet{GC08} and \citet{WBetal08}.
\citet{GC08} used the same {\it Spitzer} data of N\,44 and
similar selection criteria, but identified only 49 YSOs.
This is understandable because their automated search for point
sources was unable to find faint sources near a bright neighbor
or over a bright background.
Such faint sources can be found and measured more easily with 
human intervention, as we did in this paper for the small area
of N\,44 only.

The comparison with the YSO sample of \citet{WBetal08} is not as
straightforward because they used a very complex set of criteria.
Within the area of N\,44, they found only 19 YSO candidates,
of which one was identified as a planetary nebula (PN).
In Figure~\ref{SAGE_comp}a, we mark the locations of these YSO
candidates and our YSOs in the 8 \um\ image of N\,44, and in 
Fig.~\ref{SAGE_comp}b, we compare our sources with 
\citet{WBetal08} YSO candidates in a [8.0] vs.\ ([4.5]$-$[8.0]) CMD.
The differences can be summarized as the following: \begin{enumerate}
\item  Within the YSO wedge bounded by [4.5]$-$[8.0] $\ge$ 2.0
and [8.0] $\ge 14.0-$([4.5]$-$[8.0]), we find 60 YSOs, while
\citet{WBetal08} find only 13 YSO candidates, a subset of our sample.
To investigate why the majority of the YSOs in N\,44, including bright
sources with [8.0] $<$ 7.0, are missed in Whitney et al., we further
compare our sources with their original catalog, i.e., the SAGE point 
source catalog (kindly provided by Remy Indebetouw and Marta Sewilo).
We find that the missed YSOs do not result from different selection
criteria, but from defining point sources and applying signal-to-noise 
(S/N) thresholds when making final lists in these two studies.
For a survey of the entire LMC, the SAGE catalog used stringent 
parameters for the automated search for point sources and Whitney 
et al.\ applied a high S/N threshold.
The former discards any irregular point sources compared to IRAC PSFs.
The latter tends to exclude sources near bright neighbors or over bright 
background, as shown in Fig.~\ref{SAGE_comp}a, since varying backgrounds 
make photometry more difficult, and the SAGE pipeline increases the 
photometric uncertainties in such areas, reporting 
conservative S/N ratios.
\item Six YSO candidates of the Whitney et al.\ sample were rejected 
by our stringent color-magnitude criteria for YSO selection.  The SEDs 
of these six sources are shown in Figure~\ref{SAGE_comp}c. 
With the addition of optical fluxes in the SEDs, it can be more clearly 
seen that at least YSO candidate W546 is likely an evolved star with 
circumstellar dust and that W530 and W574 may be background galaxies.
\end{enumerate}

We conclude that our sample of massive YSOs in N\,44 is the most
complete among the three compared.

\section{Classification of Massive YSOs}

As mentioned earlier, there is not yet a well-defined classification system
for massive YSOs.
In Robitaille et al.'s (2006) study of YSO models, they suggested a physical 
classification scheme for YSOs of all masses.
This classification is analogous to the Class scheme for low-mass YSOs,
but uses physical quantities, instead of the slope of the SED, to define
the evolutionary stage of the models.
The three stages are defined by the ratio of envelope accretion rate 
($\dot{M}_{\rm env}$) to central stellar mass ($M_\star$) and the ratio
of disk mass ($M_{disk}$) to $M_\star$:
Stage I sources have $\dot{M}_{\rm env}/M_\star > 10^{-6}$~yr$^{-1}$, 
Stage II sources have $\dot{M}_{\rm env}/M_\star < 10^{-6}$~yr$^{-1}$ and 
$M_{\rm disk}/M_\star > 10^{-6}$, and Stage III sources have 
$\dot{M}_{\rm env}/M_\star < 10^{-6}$~yr$^{-1}$ and 
$M_{\rm disk}/M_\star < 10^{-6}$.
These physical quantities are not directly observable; they have to
be inferred from model fits to the observed SEDs.
This classification scheme, while physical, is model-dependent and
the comparison with observations requires data taken with sufficient 
angular resolutions to separate among individual sources and from 
backgrounds, which is not always the case for LMC objects.
Thus, it is not an appropriate description of observed properties 
of YSOs.

Here we suggest an empirical classification of the observed SEDs of
massive YSOs, and use a ``Type'' nomenclature for distinction from 
the ``Classes'' for low-mass YSOs and the ``Stages'' for model YSOs.
In our scheme: \begin{itemize} 
  \vspace{-0.3cm}
  \itemsep-0.05cm
\item[]{\bf Type I} 
 has an SED rising steeply from near-IR to 
 24 \um\ and beyond, indicating that the radiation is predominantly
 from a circumstellar envelope.
 Examined in images, a Type I YSO is not visible at optical or even 
 in the $J$ band, but emerges in the $K_s$ band and continuously 
  brightens up all the way to 24 and 70 \um.
 Type I YSOs are usually in or behind dark clouds; an example
 is given in Figure~\ref{fig:ysoimg}a, where the YSO's 
 optical and IR images, SED,  and location in the [8.0] 
 vs.\ ([4.5]$-$[8.0]) CMD are shown.
\item[]{\bf Type II} shows an SED with a low peak at optical wavelengths 
 and a high peak at 8--24 \um, corresponding respectively to the 
 central source and the inner circumstellar material (e.g., a disk)
 and indicating that the envelope has dissipated.
 In images, a Type II YSO would appear faint at optical, brighten 
 up from $J$ to 8 \um, and appear faint again at 24 \um. 
 An example of a Type II YSO is given in Fig.~\ref{fig:ysoimg}b. 
\item[]{\bf Type III} shows an SED with bright optical stellar emission 
 and  modest dust emission at near- and mid-IR, indicating that the
 young star is largely exposed but still surrounded by remnant 
 circumstellar material.
 In direct images, a Type III YSO is bright at optical, but fades
 toward longer wavelengths.
 It may even be surrounded by an \hii\ region; as described later 
 in \S 7.3, one of the Type III YSOs has an {\it HST} \ha\ 
 image available and it shows a compact \hii\ region.
 An example of a Type III YSO is given in Fig.~\ref{fig:ysoimg}c.
\end{itemize}

As our classification is based on mainly the SEDs and as
the observed SED of a somewhat evolved YSO with bipolar cavities 
is dependent on its viewing angle, an unknown fraction of our 
Type I YSOs may have the same physical structure of the Type II 
YSOs but viewed outside the cavity.
Moreover, not all YSOs can be classified unambiguously into these three
types as some YSOs may be transitional between two evolutionary stages.
A number of sources show SEDs resembling two adjacent types; for these
we simply classify them as I/II or II/III.

Additional complications arise because of the limited angular 
resolution of {\it Spitzer} and complex surroundings of YSOs.
First, IRAC's $\sim 2''$ angular resolution translates
into 0.5 pc in the LMC, which can hide a small group of YSOs.
Indeed, 20 of the YSOs are resolved by our MOSAIC or ISPI 
images into multiple sources or even small clusters within
the IRAC PSF, and another 19 YSOs appear more extended than 
the PSF of MOSAIC and ISPI.
While the ISPI $JK_s$ images allow us to identify the dominant
YSO among the multiple sources for accurate photometric measurements,
the IRAC measurements correspond to the integrated light of all 
sources, causing uncertainty in the classification, especially 
when two or more YSOs are present within the IRAC PSF.
Fig.~\ref{fig:ysoimg}d shows an example of a YSO resolved into
multiple sources in MOSAIC and ISPI images.
Second, YSOs are often found in dark clouds and dust columns.
These interstellar features can be identified in the optical images 
as dark patches against dense stellar background or photoionized
bright-rimmed dust pillars.
The emission from these dust features can be blended with
that of the YSOs, especially at the 24 \um\ band, causing large 
uncertainties in the YSO classification.
Fig.~\ref{fig:ysoimg}e shows an example of a YSO at the tip
of a dust column in N\,44F.

Our classification of the 60 YSOs and remarks on their multiplicity 
and association with dark cloud and dust column are given in 
Table~\ref{ysoclass}.
Although for multiple systems the IRAC and MIPS fluxes are 
contaminated by other sources, these fluxes are most likely
dominated by the most massive (and hence the brightest) YSO 
since the luminosity of a YSO is proportional to its mass 
cubed \citep{BM96}. 
Thus, for a YSO that appears single or is clearly the dominant source 
within the IRAC PSF, its SED can be modeled to assess its physical 
properties, as discussed in the next section.

\section{Determining YSO Properties from Model Fits of SEDs}

\subsection{Modeling the SEDs}

The observed SED of a YSO can be compared with model SEDs, and the 
best-fit models selected by the $\chi^2$ statistics can be used to
infer the probable ranges of physical parameters for the YSO.
This can be carried out with the fitting tool Online SED
Model Fitter\footnote{The Online SED Model Fitter is available at 
http://caravan.astro.wisc.edu/protostars/fitter/index.php.} 
\citep{RTetal07}.
The database of this fitting tool includes model SEDs 
pre-calculated for 20,000 YSO models each with 10 different 
viewing angles.
A typical YSO model assumes a stellar core, a flared accretion disk, and 
a rotationally flattened infalling envelope with bipolar conical 
cavities.
To fully describe this complex structure, each model is defined 
by 14 parameters \citep{RTetal06}.
The stellar parameters include the star's mass ($M_\star$), radius 
($R_\star$), and temperature ($T_\star$); these are sets of 
parameters at different ages ($t_\star$) along the pre-main-sequence
stellar evolutionary tracks modeled by \citet{BM96} and \citet{SLetal00}.
For a given set of $M_\star$, $R_\star$, and $T_\star$, the stellar radiation
determined from stellar atmosphere models of \citet{Ku93} or \citet{BH05} 
is adopted and fed to the radiative transfer code to produce the YSO's 
model SED.
The disk parameters include the disk's accretion rate ($\dot{M}_{\rm disk}$), 
mass ($M_{\rm disk}$), inner radius ($R^{\rm min}_{\rm disk}$), 
outer radius ($R^{\rm max}_{\rm disk}$), scale height factor 
($z_{\rm factor}$), and flaring angle ($\beta$).
The envelope parameters include envelope accretion rate 
($\dot{M}_{\rm env}$), envelope outer radius ($R^{\rm max}_{\rm env}$), 
cavity density ($\rho_{\rm cavity}$), cavity opening angle 
($\theta_{\rm cavity}$), and the ambient density ($\rho_{\rm ambient}$)
at which the density of the envelope reaches the lowest value as the 
density of the ambient ISM.
Outside of the envelope is the foreground extinction ($A_V$). 
The 20,000 YSO models are generated to sample each 
of the disk and envelope parameters and the ambient density
for a variety of pre-main-sequence stellar model \citep{RTetal06}.

We have used the above SED fitting tool to analyze 36 YSOs in 
our sample that appear single or are clearly the dominant source 
within the IRAC PSF.
The input parameters of the SED model fitter include the
fluxes of a YSO and uncertainties of the fluxes.
The uncertainty of a flux has two origins: the measurement itself
and the absolute flux calibration.
The fluxes and their measurement errors are given in Table~\ref{ysoclass}.
The calibration errors are 5\% in $U$, 3\% in $B$, $V$, 10\% in $I$, 
$J$, $K_s$, 3.6, 4.5, 5.8, 8.0, and 24 \um, and 20\% in 70 \um\ 
(Chen 2007; Gruendl \& Chu 2008; IRAC Data Handbook; MIPS Data Handbook).
The total uncertainty of a flux is thus the quadratic sum of the
measurement error and the calibration error.
The resulting best-fit and acceptable SED models are shown in 
Figure~\ref{fig:fit}, in which the YSOs are arranged in the order of 
Types I, I/II, II, II/III, and III (from our empirical classification 
in \S5), and within each type in order of increasing [8.0] magnitude.
In each panel, data points are plotted in filled circles and upper limits 
are plotted in filled triangles; error bars are plotted but they are
usually smaller than the plot symbols.
The best-fit model, with minimum $\chi^2$, is shown in solid black line, 
and the radiation from the stellar core reddened by the best-fit $A_V$ 
is shown in dashed black line.
For most YSOs, their SEDs can be fitted similarly well by a range of 
models, and these ``acceptable'' models are plotted in grey lines (see 
Robitaille et al.\ 2007 for definitions of $\chi^2$ and acceptable
fits).

The results of the SED model fits are given in Table~\ref{sedfits},
where the YSOs are listed in the same order as in Fig.~\ref{fig:fit}.
The source name, [8.0] magnitude, and type from our empirical classification
are listed to the left of the table, and physical parameters of the 
best-fit models are listed to the right.
Among the 14 parameters of each model, we have only listed the stellar
parameters ($M_\star$, $T_\star$, $R_\star$, $t_\star$), accretion rates
($\dot{M}_{\rm env}$, $\dot{M}_{\rm disk}$), disk mass ($M_{\rm disk}$),
and foreground extinction ($A_V$).
We have also listed the viewing angle ($i$, the angle between the 
sightline and the polar axis), and the total luminosity ($L_{\rm tot}$).
In addition to the parameters of the best-fit model, we have also 
used the acceptable models to show a possible range of stellar mass
($M_\star$ Range), and used $\dot{M}_{\rm env}/M_\star$ and
$M_{\rm disk}/M_\star$ from the acceptable models to estimate
a possible range of its evolutionary stage, the Stage Range, as defined 
by \citet{RTetal06}.
The possible ranges of stellar mass and evolutionary stage are also
given in Table~\ref{sedfits}.

The SED fits of the 36 YSOs, as displayed in Fig.~\ref{fig:fit}, 
show different degrees of goodness-of-fit among our empirically
defined YSO types.
The seven Type I and I/II YSOs have model SEDs agree well with 
the observed SEDs, although the best and acceptable models span
a large mass range.
The 21 Type II and II/III YSOs show good agreement between
model and observed SEDs, except at 4.5 \um, where the observed
fluxes are systematically lower than the modeled. 
This discrepancy will be discussed below in \S6.2.1.
Among the eight Type III YSOs analyzed, only three of them show good
agreement in the SED fits, while the other five exhibit significant 
discrepancies between model and observed SEDs.
These discrepancies and possible causes will be discussed in \S6.2.2.

\subsection{Significant Discrepancies Between Model and Observed SEDs}

\subsubsection{PAH Emission in Massive YSOs}

A great majority of our Type II and III YSOs show a brightness dip at 
4.5 \um\ in their SEDs (Fig.~\ref{fig:fit}).
This dip is not an absorption feature as it appears.  
Instead, it is caused by PAH emission features at 3.3, 6.2, 7.7, 
and 8.6 \um\ in the other three IRAC bands \citep{LD01,LD02,DL07}.
The emission mechanism of these features is ultraviolet (UV) excitation
of PAHs followed by IR fluorescence \citep{ATB89}; thus these
emission features have been observed in regions 
with UV radiation and dust, such as disks around Herbig Ae/Be stars 
\citep[e.g.,][]{RB03,VBetal04} and photodissociation regions (PDRs) 
surrounding \hii\ regions \citep[e.g.,][]{HT97}.
Our Type II and III YSOs have similar physical conditions, and the 
resultant PAH emission cause the apparent dip at 4.5 \um\ in the SEDs.

The PAH emission of the YSOs is unlikely to originate from a 
large-scale, diffuse interstellar structure, e.g., a superbubble 
rim, since the contamination from such interstellar emission is 
minimized by background subtraction in the photometric measurement.
Furthermore, there is a correlation between PAH emission and our YSO 
types: PAH emission is not observed in Type I, appears in some of 
the Type I/II YSOs, and is almost ubiquitous in Type II and III YSOs.
This correlation argues against the large-scale interstellar origin.
Therefore, the PAH emission in our YSOs is most likely of a 
circumstellar origin, such as a disk or a compact \hii\ region, or both.
A Type I YSO is deeply embedded in a massive envelope, of which only
the dust continuum is observable, and thus no PAH emission is expected.
A Type II YSO may show prominent PAH emission from the disk, if viewed
through its bipolar cavities.
A Type III massive YSO will ionize its surrounding gas and show
PAH emission from its PDRs.
While the SEDs reveal signatures of PAH emission, the geometry of 
the emitting region (whether a disk or a PDR of an \hii\ region) 
cannot be determined without more detailed spectral information.

The SEDs of massive YSOs show evidence of PAH emission, but the
SED models of \citet{RTetal06} do not include PAHs or small grains.
This fundamental difference causes the dust continuum determined 
from SED fitting to be biased by the three IRAC bands that contain 
PAH emission.
Consequently, the observed 4.5 \um\ fluxes are below the model fits.
Moreover, the error in the best-fit continuum level is propagated 
to the disk parameters, which are predominantly determined from
the fluxes in the IRAC bands.
Thus, SED model fits for YSOs with PAH emission may have large 
errors in the viewing angle or disk mass.
These errors may be further propagated into the simultaneously 
fitted envelope and stellar parameters, but the effects are 
unlikely to be large because the stellar and envelope parameters 
are determined mainly by the SED at optical and mid-IR wavelengths,
respectively.

\subsubsection{Discrepancies in Type III YSOs}

Among the eight Type III YSOs, five show significant discrepancies 
between the best-fit models and the observed SEDs: 052129.7$-$675106.9,
052315.1$-$680017.0, 052157.0$-$675700.1, 052159.6$-$675721.7, and
052340.6$-$680528.5 (Fig.~\ref{fig:fit}).
The brightest object, 052129.7$-$675106.9, is peculiar and will be 
discussed later in \S7.2.
The other four YSOs show deep V- or U-shape SEDs with the flex
point in the IRAC bands, where the best-fit model fluxes are 
much higher than the observed fluxes, exhibiting the most prominent
discrepancies.

To understand these discrepancies, we can divide the SED of a YSO 
into three segments, optical/near-IR ($UBVIJK_s$), near/mid-IR 
(IRAC bands), and mid/far-IR (MIPS bands and beyond), and look into 
the dominant origin of emission for each segment.
At the shortest wavelengths, the optical/near-IR segment,
the radiation is dominated by stellar photospheric emission;
for example, Figure~\ref{fig:sfit} shows the optical/near-IR
segment of the SEDs of four Type III YSOs can be well fitted
by stellar atmosphere models \citep{Ku93}.
At longer wavelengths, the stellar emission diminishes and
dust emission rises. 
The near/mid-IR segment consists of emission from PAHs and warm
dust, while the mid/far-IR segment consists of emission mostly 
from colder dust.
Qualitatively, warm dust that emits at near/mid-IR wavelengths 
needs to be near the star in order to reach the required emitting
temperatures; therefore, the near/mid-IR segment of a YSO's SED is 
dominated by disk emission.
Colder dust that emits at mid/far-IR wavelengths is farther
away from the star, and thus the mid/far-IR segment of a YSO's SED
is dominated by envelope emission.

The aforementioned discrepancies between observations and best-fit
models of SEDs of Type III YSOs in the IRAC bands suggest that the
YSOs do not have as much disk emission as in the models.
Below we analyze the SEDs quantitatively to understand their
physical implications.
The mid- to far-IR parts of the V- or U-shaped SEDs of these 
Type III YSOs (Fig.~\ref{fig:fit}) suggest that the dust continuum
peaks at $\ge$24 \um.
The corresponding blackbody temperature of the dust is thus $\le$120 K.
The distance of such dust temperature to the central star can be estimated
if the stellar spectral type and the implied stellar effective 
temperature are known.
Among these four YSOs, 052157.0$-$675700.1, 052159.6$-$675721.7, 
and 052340.6$-$680528.5 have been spectroscopically classified as 
O8.5V$+$neb, O7.5V$+$neb, and B[e] stars, respectively \citep{OM95}, 
and 052315.1$-$680017.0 has $(U-B)=-0.79$, $(B-V)=0.09$, and $V=14.10$ 
(Table~\ref{ysoclass}) that are consistent with the colors and magnitudes
of a B0-O5 V  star \citep{SK82} reddened by $A_V \sim 1.2$.
For a dust temperature of 120 K and an albedo of 0.5, the closest
distance of such dust to a B0 V star with a stellar effective
temperature of 30,000 K is 860 AU; this distance is larger for 
stars with higher effective temperatures.
This result implies a low dust content, or a hole, within the central
$\sim$1000 AU.

While our assessment of the observed SEDs suggests 
$R^{\rm min}_{\rm disk}$ $>$ 860 AU, no models with 
$R^{\rm min}_{\rm disk}$ $>$ 100 AU are available in the 
database of pre-calculated SEDs for model fitting \citep{RTetal07}.
To illustrate how $R^{\rm min}_{\rm disk}$ affects the SEDs, we
have used the \citet{WBetal03} radiative transfer 
code\footnote{The code is available at 
http://caravan.astro.wisc.edu/protostars/codes/index.php.}
to calculate SEDs for a YSO model with stellar parameters
appropriate for the observed spectral types but with customized
disk parameters.
The YSO model \#3019532 is chosen because its $T_\star=$ 34718 K 
and $R_\star=$ 5.628  $R_\odot$ are representative of late-type O stars.
In this Stage III YSO model, the envelope has dissipated completely.
We have calculated two SEDs with disk radii of 50--1000 AU and 
1000--2000 AU, respectively; these model SEDs are overplotted on
the observed SED of 052157.0$-$675700.1 in Figure~\ref{fig:fit_hole}.
The model SED with $R^{\rm min}_{\rm disk}$ = 50 AU show prominent
excess emission in the near/mid-IR bands, while the model 
SED with $R^{\rm min}_{\rm disk}$ = 1000 AU adequately reproduce the
V-shaped SED without excess near/mid-IR emission.
This comparison suggests that dust has been cleared out to 
nearly 1000 AU.
This clearing of dust is consistent with the expectation from
YSOs at late evolutionary stages.
We thus conclude that the large discrepancies between the best-fit
models and the observed SEDs are caused by the absence of appropriate 
model SEDs in the database of the Online SED Model Fitter.

Similar V-shaped SEDs have been observed in Herbig Ae/Be stars RCW34 
and TY Cra, and the SED modeling suggested an analogous evolutionary 
stage \citep{HLetal92}.
Similar near-IR dips have also been observed in SEDs of the low-mass 
YSOs DM Tau and GM Aur; these dips have been interpreted as the clearing
of dust from the inner regions of their disks \citep{RWetal03,CNetal05}.

\subsection{Evolutionary Stage of YSOs}

\subsubsection{Type vs. Stage}

We use our analysis of 36 YSOs to compare our empirical classification, 
Type, with the theoretical classification, Stage \citep{RTetal06}.
For each YSO, the $\dot{M}_{\rm env}/M_\star$ and $M_{disk}/M_\star$ ratios
from the best-fit and acceptable models are used to determine its Stages,
and the possible range of Stage is listed in Table~\ref{sedfits} under
the column ``Stage Range''.
At first glance, there does not seem to be any systematic correspondence
between Types and Stages.

The lack of overwhelming correspondence may be attributed to three reasons.
First, the definitions of Stages are somewhat arbitrary, and it is not
even known whether the Stages can be universally applied to YSOs of all
masses.  Furthermore, a Stage II YSO viewed along directions near the
disk plane will mimic a Stage I YSO in the SED, causing ambiguity.
Second, the 24 \um\ fluxes are vital in constraining the model fits,
but they are the most uncertain measurements, especially for the faint
YSOs.  For example, the uncertain 24 \um\ fluxes of Type I YSOs 
052216.9$-$680403.6 and 052211.9$-$675818.1 cause their model fits
to be ill constrained (see Fig.~\ref{fig:fit}).
Third, the limited angular resolution of {\it Spitzer} may cause 
inclusion of extraneous dust emission from unresolved \hii\ regions.
This is particularly relevant to YSOs at late evolutionary stages.
For example, the Type III YSO 052207.3$-$675819.9, as shown in
\S7.3, has a small \hii\ region that is resolved only in {\it HST}
WFPC2 images; without these WFPC2 images, circumstellar disk or
envelope would have been invoked to explain the YSO's far-IR emission.

Still, we expect the extreme Stage I and III YSOs to correspond to
our Type I and III, respectively, since these extreme types are
either deeply embedded in massive envelopes or have little 
circumstellar material so that their SEDs are less dependent 
on the viewing angles.
Examined closely, it can be seen that Type I YSOs with good flux
measurements at 24 \um\ are indeed well fitted by models at Stage I,
and Type III YSOs without unresolved \hii\ regions indeed correspond
to Stage III, if they have good 24 \um\ flux measurements and the
model fits do not have near/mid-IR excesses (see \S6.2.2).
In summary, we conclude that meaningful comparisons between Types 
and Stages can be made only if good mid/far-IR flux measurements
and high angular resolution images are available.

\subsubsection{Evolutionary Stages of YSOs in Ultra-compact \hii\ Regions}

Some young massive stars are known to be associated with small
(diameter $\le10^{17}$ cm), dense ($\ge10^4$ cm$^{-3}$) regions of
ionized gas, called ultracompact \hii\ regions (UCHIIs).
Five UCHIIs have been identified in N\,44: B0522$-$5800,
B0523$-$6806(NE), B0523$-$6806(SE), B0523$-$6806(SW), and
B0523$-$6806 \citep{IJC04}.
All five UCHIIs are coincident with {\it Spitzer} sources within
1$''$.
The {\it Spitzer} counterpart of B0523$-$6806(SE), 052324.8$-$680641.6, 
is faint and falls below the cutoff line of [8.0] 
$\ge 14.0-$([4.5]$-$[8.0]) in the [8.0] vs.\ [4.5]$-$[8.0] CMD, 
where background galaxies and YSOs with masses $\le4~M_\odot$ are
located.
As low-mass stars cannot produce UCHIIs, this source
is most likely a background star-forming galaxy.
We have examined the MOSAIC $I$, ISPI $JK_s$, and {\it Spitzer} 
IRAC and MIPS images of this source.  The $I$ and $J$ images show
a brighter source and a faint source separated by less than 1$''$.
The faint source becomes the brighter of the two in the $K_s$
band, and is the dominant source in the {\it Spitzer} bands.
This source is not in a giant molecular cloud or surrounded
by bright diffuse PAH emission.  These results support the
identification of B0523$-$6806(SE), or 052324.8$-$680641.6, 
as a background star-forming galaxy.

The {\it Spitzer} counterparts of the other four UCHIIs are bright 
and have been identified as Type I-II YSOs (see Table~\ref{uch2}).
As these sources are among the top 10 most luminous
YSOs in N\,44 at 8.0 \um, they have good 24 \um\ flux measurements 
and consequently models of their SEDs are well constrained.
The stellar masses determined from the best-fit models 
to their SEDs can be translated into spectral types, assuming
the relationship for main sequence stars.
These spectral types can be compared with those implied by the 
ionizing fluxes determined from radio continuum observations 
\citep{IJC04}.
As seen in Table~\ref{uch2}, excellent agreement exists between 
these two independently determined spectral types.  

The development of a UCHII depends on not only the ionizing flux
provided by the central star, but also the opacity of the 
circumstellar medium.
For infalling rates higher than some critical value, $\dot{M}_{\rm crit}$,
the circumstellar medium will have such high opacities that
the ionized region will be too small and too optically thick to
be detectable \citep{CE02}.
For the respective spectral types of the central stars, the 
$\dot{M}_{\rm crit}$ of the four UCHIIs in N\,44 are computed to 
be $\sim 1-4\times10^{-5} M_\odot$~yr$^{-1}$ (Table~\ref{uch2}).
Compared to the $\dot{M}_{\rm env}$ determined from the best-fit 
models for these four YSOs (Table~\ref{uch2}), it is seen that
only the Type II YSO 052255.2$-$680409.5 has 
$\dot{M}_{\rm env}~< ~\dot{M}_{\rm crit}$, and the other three YSOs
have $\dot{M}_{\rm env}~\gg ~\dot{M}_{\rm crit}$.
We have further examined whether any acceptable models of the latter
three YSOs yield smaller envelope accretion rates.
We find that only 052343.6$-$680034.2 has a few models with 
$\dot{M}_{\rm env} \le \dot{M}_{\rm crit}$, but these models have
 $T_\star =$ 38000--40000 K, higher that that for an O9 V star 
($T_\star \sim$ 33000 K) estimated from the ionization 
requirement of the UCHII.
All the other acceptable models have $\dot{M}_{\rm env} \gg
\dot{M}_{\rm crit}$. 

Our limited sample shows that the Type II YSO in UCHII has 
$\dot{M}_{\rm env} < \dot{M}_{\rm crit}$, and the Type I and I/II 
YSOs in UCHIIs have $\dot{M}_{\rm env} \gg \dot{M}_{\rm crit}$.
As Types I and I/II YSOs are still dominated by envelopes,
it is possible that most of the infalling envelope material 
is used in forming an accretion disk, instead of the stellar 
core, as modeled by \citet{YS02}.
Therefore, $\dot{M}_{\rm env}$ should not be interpreted as
the accretion rate of mass onto the stellar core.

\subsection{Masses of YSOs}

In our sample of 60 YSOs in N\,44, 24 do not have mass estimates
from SED model fits as their IRAC fluxes are contaminated by
neighboring stars; nevertheless, it can be judged from 
their lower brightnesses that they are probably on the low-mass 
end in the sample.
The other 36 YSOs have reliable SEDs that can be modeled to determine 
their masses; their mass estimates from the best and acceptable fits 
are listed in Table~\ref{sedfits}.
Although many of these YSOs have large uncertainties in their mass 
estimates, i.e., large $M_\star$ Range, 30 of them have $M_\star$ Range all
greater than $8 M_\odot$ and are thus most likely bona fide massive YSOs.
The remaining six YSOs have $M_\star$ Range extending from intermediate- 
to high-mass; these may also be massive YSOs, particularly for the four 
with best-fit masses $\ge 8 M_\odot$.

It has been suggested that the criterion [8.0] $\le 8.0$ may be used
to select massive YSOs in the LMC \citep{GC08}.
We find that indeed the $M_\star$ Ranges of the brightest YSOs, with
[8.0] $\le 8.0$, are all $\ge 8 M_\odot$.
This criterion may be too conservative, as almost all (25 out of 27)
YSOs with [8.0] $\le 9.0$ still have masses $\ge 8 M_\odot$

At the high-mass end, nine YSOs have $M_\star$ Ranges all 
$\ge 17 M_\odot$; these are most likely O-type YSOs.
Five YSOs that show $M_\star$ Range with an upper mass limit 
$\ge 17 M_\odot$ and a lower mass limit $\ll 17 M_\odot$;
these may or may not be O-type stars.
To improve the census of O-type YSOs, we have checked the 
optical spectral classifications of the Type III YSOs to search 
for O stars that are missed by our estimates of masses from 
SED model fits.
We find that two Type III YSOs with optical spectral types of
O7.5\,V and O8.5\,V \citep{OM95} have masses determined from the 
SED model fits to correspond to B0--3\,V spectral types
(052157.0$-$675700.1 and 052159.6$-$675721.7).
Therefore, there exist at least 11 O-type YSOs in N\,44.
These most massive YSOs will be discussed further in \S7.1.

\section{Massive Star Formation in N\,44}

It is difficult to study the relationship between interstellar
conditions and the formation of massive stars because massive stars'
UV radiation fluxes and fast stellar winds quickly ionize and
disperse the ambient ISM.
Massive YSOs, on the other hand, have not significantly altered the
physical conditions of their surrounding medium on a large scale, and 
thus can be used to probe massive star formation.
The large number of massive YSOs found in N\,44 provides an excellent 
opportunity to investigate issues such as relationship between
star formation properties and interstellar conditions, progression
of star formation, and evidence of triggered star formation.

\subsection{Interstellar Environments and Star Formation Properties}

We examine the star formation properties of the molecular clouds 
in N\,44 as these clouds contain the bulk material to form stars.
The NANTEN CO survey of the LMC \citep{FYetal01} shows three
large concentrations of molecular material in N\,44, i.e.,
the central, southern, and northern peaks (Figure~\ref{fig:yso_pos}).
These three concentrations exhibit different numbers of massive stars
formed in the last few Myr, as evidenced by their different amounts
of ionized gas.
The central molecular peak is associated with prominent \ha\
emission from a supershell and bright \hii\ regions along the shell rim.
Star formation has been occurring at this site for an extended period 
of time, with the supershell encompassing 10-Myr old massive stars and the 
bright \hii\ regions containing $\sim$5-Myr old massive stars \citep{OM95}.
The southern molecular peak shows one bright \hii\ region and 
several smaller, disjoint \hii\ regions.
The massive stars of these \hii\ regions have not been studied 
spectroscopically, but the absence of shell structures produced 
by fast stellar winds and supernovae implies that massive 
stars are most likely also $\sim$5-Myr old or that star 
formation has started only in the last few Myr.
The northern molecular peak has a couple small \hii\ regions,
indicating that only modest massive star formation has taken place.

Fig.~\ref{fig:yso_pos} shows that almost all of the YSOs in N\,44 
are found in molecular clouds and about 75\% of the YSOs are 
congregated toward the three molecular peaks.
The YSOs of the three molecular peaks show different characteristics 
in their spatial distributions and interstellar environments.
The central molecular peak has the highest concentration of YSOs, as 
21 of them aggregate in the prominent \hii\ regions along the 
southwest rim of the supershell.
The southern molecular peak has loosely distributed YSOs,
and most of these 11 YSOs are associated with the
disjoint \hii\ regions.
The northern molecular peak has 12 YSOs that are also loosely distributed,
but the majority of the YSOs are not associated with any ionized gas.

The YSOs of the three molecular peaks in N\,44 also show differences
in their mass distributions.
In Fig.~\ref{fig:yso_pos}, we have marked the YSOs with circles in
three sizes that represent O-type stars with $M_\star \ge$ 17 $M_\odot$,
B-type stars with $M_\star \ge$ 8 $M_\odot$, and intermediate-mass
stars with $M_\star <$ 8 $M_\odot$, respectively.
It is striking that $\sim$80\% of the O-type and B-type YSOs are 
in or adjacent to \hii\ regions.
The YSOs with intermediate masses or without mass estimates do not
show such strong bias in spatial distribution, although this may 
be partly due to an observational bias, as these YSOs are fainter 
and it is more difficult to detect faint YSOs over the bright
background dust emission in \hii\ regions.
The central molecular peak, having the most prominent \hii\
regions, possesses the highest concentration of O- and B-type YSOs.
The southern molecular peak has a respectable number of O-
and B-type YSOs, while the northern molecular peak has no O-type
YSOs at all. 

The characteristics of the current star formation in the three 
molecular peaks of N\,44 appear to be dependent on the massive 
star formation that occurred in the recent past.
It is possible that the pattern of star formation is controlled
by properties of the molecular clouds.
The central, southern, and northern peaks correspond to the giant
molecular clouds (GMCs) LMC$/$M5221$-$6802, LMC$/$M5239$-$6802, and 
LMC$/$M5221$-$6750 cataloged by \citet{MNetal01}.
Their FWHM line-widths ($\Delta V$) at the peak position are 7.2, 
15.8, and 3.8 km~s$^{-1}$, respectively.
It is interesting to note that the $\Delta V$ of the southern 
molecular peak of N\,44 is the highest and that of the northern 
molecular peak is nearly the lowest among all GMCs in the LMC.
While the small $\Delta V$ of the northern molecular peak reflects
the low level of stellar energy feedback in the last few Myr,
the larger $\Delta V$ of the other two peaks do not seem to scale 
with star formation activity.
Detailed mapping of these three molecular clouds is needed to
search for fundamental differences among these three clouds
that might be responsible for their different star formation
characteristics.

\subsection{{\it HST} Images: a Closer Look at the YSOs}

To examine the immediate surroundings of the YSOs in N\,44, we have 
searched the {\it HST} archive and found useful WFPC2 images of 
three fields that contain YSOs.
These three fields encompass the \hii\ regions N\,44C, N\,44F, and 
N\,44H, respectively, and their locations in N\,44 are shown in
Fig.~\ref{fig:n44opt}a.
These \hii\ regions and their associated YSOs and ionizing stars 
are individually discussed below.

N\,44C is a bright \hii\ region located at the southwest rim
of the N\,44 superbubble.
The {\it HST} \ha\ image of N\,44C shows an overall shell morphology:
the northern part is bright and centered on the ionizing O7 V star
\citep{OM95}, while the southern part is faint and consists of multiple
circular filaments superposed by radial filaments streaming to the
southwest (Figure~\ref{fig:n44c_img}a).
It is not clear whether these southern radial and circular filaments
are physically associated with each other.
N\,44C abuts against a dark cloud to the north,
which coincides with the core of the GMC M5221$-$6802 revealed
by high-resolution ESO-SEST observations \citep{CYetal97}.
The observed density variations suggest that N\,44C is a blister 
\hii\ region on the surface of a molecular cloud.

Seven YSOs are within the field-of-view of the {\it HST} \ha\
image of N\,44C.  
One is projected at the base of a bright-rimmed dust pillar in 
the northwest part of N\,44C. 
Four are in the dark cloud 2--3 pc exterior to the northwest 
edge of N\,44C; all four are superposed on diffuse 8 \um\ PAH 
emission (Fig.~\ref{fig:n44c_img}b), indicating that they may be 
associated with PDRs.
One is located further northwest and projected in the N\,44 supershell 
rim; it is also superposed on large-scale diffuse 8 \um\ PAH emission 
and may be associated with PDRs.
The last one is located in the southwest outskirts of N\,44C; it 
does not have prominent PDRs around it.
It is remarkable that the majority of the YSOs associated with N\,44C
are in a molecular core and superposed on prominent PDRs.
Apparently they were formed in an interstellar environment that
has been affected by energy feedback from stars that were formed
a few Myr earlier.

N\,44F is a bright, ring-shaped \hii\ region located at the 
northwest outskirts of the superbubble.
The \hii\ region is ionized by an O8III star \citep{WBD97}.
Two prominent bright-rimmed dust pillars can be readily recognized 
in the WFPC2 \ha\ image (Figure~\ref{fig:n44f_img}a).
One of these dust pillars has a YSO, 052136.0$-$675443.4,
emerging at its tip, reminiscent of those seen in the Eagle Nebula
\citep[M\,16,][]{HJetal96}.
The mass estimated from SED fits for this YSO is $\sim 6-11~M_\odot$,
more massive than those with $\sim 3-4~M_\odot$ 
seen in the Eagle Nebula \citep{TSH02}.

N\,44H is located to the southeast of the N\,44 superbubble.
This \hii\ region contains the luminous blue variable (LBV)
HD\,269445 \citep{SOetal84,HD94} and four blue stars whose
photometric measurements suggest spectral types of early-B
\citep{Ch07}; thus, the LBV is most likely responsible for 
photoionizing the \hii\ region.
The YSO 052249.2$-$680129.0 is projected within N\,44H toward
its bright northwest rim, $\sim 60''$, or 15 pc, from the LBV.
The WFPC2 H$\alpha$ image of this YSO shows a close pair of 
sources and another source at $\sim$3$''$ south
(Figure~\ref{fig:n44h_img}a).
The 8 \um\ image (Figure~\ref{fig:n44h_img}b) shows a bright
source and a faint extension to the south that are
coincident with the optical pair and southern source,
respectively.
It is possible that all three sources are YSOs.

\subsection{Triggered Star Formation}
 
We use the YSOs and their interstellar environment to investigate 
whether some of the current star formation in N\,44 is triggered.
Fig.~\ref{fig:yso_pos} shows that on a large scale the central 
supershell of N\,44 exhibits the most prominent association
between stellar energy feedback and star formation.
The alignment of massive YSOs in the southwest rim of the supershell
suggests that the expansion of the supershell into the molecular 
cloud has triggered the star formation.
This triggering mechanism may have been going on for $\sim$ 5 Myr
and caused the formation of the massive stars in the bright
\hii\ regions along the shell rim.

Examined closely, the majority of massive YSOs are located at 
the edges of \hii\ regions or in PDRs of dark clouds.
These are suggestive examples of star formation triggered 
by external thermal pressure raised by photoionization or
photodissociation.
YSOs are also found in bright-rimmed dust pillars.
Among them, the YSO in N\,44F, as aforementioned in \S7.2,
is in a simple interstellar structure and its immediate surroundings 
can be examined with WFPC2 images (Fig.~\ref{fig:n44f_img}a),
providing an opportunity to investigate whether the star formation 
is triggered.
For this YSO to be formed from triggering instead of merely being 
exposed by the ionization front of the \hii\ region, the formation 
time scale of the YSO should not be longer than the time for the 
ionization front to traverse from the tip to the bottom of the pillar.
For the pillar's projected length of $\sim$ 1.25 pc and a shell 
expansion velocity of 12 km~s$^{-1}$ \citep{NYetal02}, the traverse 
time is $\sim$ 0.1 Myr.
This time scale is comparable to the formation time scale for a 
10$M_\odot$ YSO, $\sim$ 0.1 Myr \citep{BM94}, making it a plausible 
case of triggered star formation.
To further determine statistically the importance of triggering
for star formation, a systematic survey of the immediate surroundings 
of YSOs with high-resolution images, such as {\it HST} images, 
is needed.

The supershell of N\,44 shows a breakout at the south rim,
and hot gas has been escaping and deflected to the east.
The juxtaposition of this hot gas flow and the star forming
regions associated with LH49 in the southern GMC is 
intriguing and may suggest triggered star formation.
However, there is no direct physical evidence that the hot
gas outflow compresses the molecular cloud to form stars.
In fact, the thermal energy of the hot gas outflow, 
$\sim 1\times10^{50}$ ergs \citep{MEetal96}, is much smaller
than the kinetic energy of the southern GMC, $\sim 3\times10^{51}$ 
ergs estimated from a mass of 2.1$\times10^{6}$ $M_\odot$ and a 
velocity dispersion of 12.6 km~s$^{-1}$ \citep{MNetal01}. 
Furthermore, the star formation in the southern GMC
is mostly concentrated in regions not in contact with the 
hot gas flow.
We conclude that the breakout of N\,44's supershell is not
responsible for triggering the star formation in the
southern molecular concentration.

\subsection{A Small \hii\ Region around YSO 052207.3$-$675819.9}

It is interesting to note that a small nebula is detected 
around YSO 052207.3$-$675819.9 projected in the dark cloud
northwest of N\,44C.
A closeup of this nebula in Figure~\ref{fig:n44c_hii} shows 
that it is detected in the \ha\ emission-line image but not in the $y$ 
continuum image, indicating that it is an \hii\ region, instead
of a reflection nebula.
This small \hii\ region has a size of 2\farcs2$\times$1\farcs5, or 
0.53pc$\times0.36$pc, and an average H$\alpha$ surface
brightness\footnote{Owing 
to the $\sim300$ \kms\ redshift of the LMC, the filter transmission of 
the red-shifted \ha\ line is $\sim93$\% of the peak transmission, thus 
the extracted \ha\ surface brightness and flux are multiplied by a 
correction factor of 1.07.} of $1.9\times10^{-14}$ 
ergs~cm$^{-2}$~s$^{-1}$~arcsec$^{-2}$, corresponding to an emission 
measure of $9.4\times10^3$ cm$^{-6}$~pc.
Adopting an average diameter of 0.45 pc as the path length of 
H$\alpha$ emission, the rms electron density of the \hii\ region
is then $\sim 145$ cm$^{-3}$.
This \hii\ region has a size comparable to those of compact \hii\ 
regions, i.e., $\lesssim 0.5$ pc, but its emission measure and 
rms electron density are much lower than the typical values for 
compact \hii\ regions, $\gtrsim 10^7$ cm$^{-6}$~pc and 
$\gtrsim 5\times10^3$ cm$^{-3}$ \citep{Fr00}.
The lower densities of this small \hii\ region suggest that it has
expanded and is more evolved than a compact \hii\ region.

This small \hii\ region provides an opportunity for us to use
its ionization requirement to assess the spectral type of the
ionizing star and compare its implied mass with that determined
from the YSO SED fitting.
The $UBV$ photometry of YSO 052207.3$-$675819.9, given in 
Table~\ref{ysoclass}, suggests that its central star is a $\sim$ 
B2 V star reddened by $A_V \sim 1.3$.
Applying this extinction correction to the observed \ha\ flux, 
we obtained an \ha\ luminosity of $L_{{\rm H}\alpha}$ =
3.7$\times$10$^{34}$ ergs~s$^{-1}$, which requires an ionizing power 
of $Q({\rm H}^0) = 7.4\times10^{11}~L_{{\rm H}\alpha}$ photons~s$^{-1}$
= $2.7\times10^{46}$ photons~s$^{-1}$.
This ionizing power can be provided by a main-sequence star of 
effective temperature of 26,200 K \citep{Pa73}, corresponding to 
a spectral type of B1 V \citep{SK82}\footnote{Note that for the 
same spectral type, \citet{Pa73} used a lower temperature scale than 
\citet{SK82}.  We have adopted Schmidt-Kaler's (1982) $UBV$ photometric
calibrations for spectral types; therefore, we use Panagia's (1973)
calibration to convert ionizing power to stellar effective temperature 
and use Schmidt-Kaler's (1982) calibration to convert from stellar 
effective temperature to spectral type.}.
The mass estimated from SED model fits for this YSO is 9--15 $M_\odot$,
corresponding to spectral type B1-2 V.
It is satisfactory that the three independent methods have produced 
consistent mass estimates.

This YSO in a small \hii\ region illustrates a caveat in analyzing
objects that are not fully resolved by {\it Spitzer} images.
The SED model fits of YSO 052207.3$-$675819.9 infers that it is
at evolutionary Stage II.
The presence of a visible \hii\ region and the visibility of
the YSO in $UBV$ bands indicate that the star has little 
circumstellar dust and that the YSO is likely at Stage III.
This discrepancy in evolutionary stage is most likely caused
by the inclusion of dust emission from the \hii\ region in 
the flux measurements, since it is not resolved in the
ISPI and {\it Spitzer} images.
It is thus important to use high-resolution images to 
examine the immediate surroundings of massive YSOs in order
not to be mis-led by the SEDs.

\subsection{The Source in N\,44A: Young or Evolved?}

The source 052129.7$-$675106.9 is coincident with a small compact
\hii\ region, identified as an \ha\ knot and cataloged as N\,44A
by \citet{He56}.
Previously, this IR source was identified as IRAS 05216$-$6753 and 
classified as an \hii\ region \citep{RPetal87} or an obscured 
supergiant or AGB \citep{WPetal92,LCetal97}.
Figure~\ref{fig:agb2yso} shows that the central star of this small
\hii\ region is bright in optical wavelengths.
\citet{Ch07} measured $V = 14.17 \pm 0.02$, 
($B-V$) $= 0.79 \pm 0.03$ and ($U-B$) $= -0.64 \pm 0.06$.
These magnitudes and colors cannot be reproduced by any combination
of normal spectral type and luminosity class. 
The observed ($U-B$) indicates a blue star, so the intrinsic
color ($B-V$)$_0$ must be in the range of 0 to $-$0.3; therefore,
E($B-V$) = 0.8--1.1, or $A_v$ = 2.5--3.4, and  $M_V$ = $-$6.8 to $-$8,
suggesting that this star is a blue supergiant of luminosity
class I.
It cannot be an AGB star.

We can make another independent estimate of the spectral type
based on the \ha\ luminosity of N\,44A.  Using the flux-calibrated
MCELS images of N\,44, we subtracted the red continuum from the 
\ha\ image and measured a continuum-subtracted \ha\ flux of 
4.7$\times$10$^{-12}$ ergs~s$^{-1}$~cm$^{-2}$ from N\,44A.
To achieve feasible ionizing power, the star has to be blue
with ($B-V$)$_0$ $\sim$ $-$0.3, thus we adopt an extinction 
of $A_v \sim 3.4$ and find an \ha\ luminosity of 
$1.4\times10^{37}$ ergs~s$^{-1}$.  This \ha\ luminosity requires 
the ionizing power of an O9\,I star.

We thus suggest that the central star of N\,44A is an O9\,I star.
The strong mid-IR excess of this star indicates the existence
of abundant dust around the star.
The anomalous combination of ($U-B$) and ($B-V$) colors may be
caused by the inclusion of extra scattered light in the $U$ band.
If the star is a young star, the dust and \hii\ region would both 
be interstellar.
If the star is an evolved star, the dust and \hii\ region would
both be ejected stellar material and show enhanced elemental 
abundances.
The current data do not allow us to distinguish whether the 
central star of N\,44A is a YSO or an evolved massive star,
such as an LBV.
High-resolution optical spectra of the \ha\ and [\ion{N}{2}] 
lines are needed to search for N-enriched ionized gas.
If the ionized gas is enriched, the central star of N\,44A
is an evolved star.
If the ionized gas is not enriched, the central star of N\,44A
must be young, otherwise its stellar wind would have blown
away the small \hii\ region.

\section{Summary}

We have observed the starburst region N\,44 in the LMC with the
{\it Spitzer} IRAC and MIPS at 3.6, 4.5, 5.8, 8.0, 24, 70, and 
160 \um, CTIO Blanco 4 m telescope ISPI in $J$ and $K_s$ and 
MOSAIC in SDSS $u$ and Johnson-Cousins $BVI$ bands.
Point sources were identified and their photometric measurements 
were made.
To identify YSOs, we first constructed an [8.0] vs. [4.5]$-$[8.0] 
CMD and used the criteria [4.5]$-$[8.0] $\ge$ 2.0 to exclude normal
and evolved stars and [8.0] $<$ 14.0$-$([4.5]$-$[8.0]) to exclude 
background galaxies.  A total of $\sim$100 YSO candidates were 
identified.  We then inspected the SED and close-up images of each 
YSO candidate in H$\alpha$, $BVIJK_s$, IRAC, and MIPS bands 
simultaneously to further identify and exclude evolved stars, 
galaxies, and dust clumps, resulting a final sample of 60 YSO 
candidates that are most likely bona fide YSOs of high or
intermediate masses.

Following the classification of Classes I, II, and III for
low-mass YSOs, we suggest the classification of Types I, II, 
and III for higher-mass YSOs, according to their SEDs.  
Type I YSOs are not detectable in optical to $J$ bands, and 
have SEDs rising beyond 24 \um.  Type II YSOs become visible 
in optical to $J$ bands, and have SEDs peaking in the mid-MIR
but falling beyond 24 \um.  Type III YSOs are bright in optical,
but still show excess IR emission.

To assess the physical properties of the YSOs, we use the Online
SED Model Fitter \citep{RTetal07} for 36 YSOs whose SEDs are
reliably determined.
We find the SEDs of  Type I and Type I/II YSOs can be modeled
well, but the SEDs of Type II and Type II/III YSOs show 
prominent discrepancies at 4.5 \um\ because the SED models 
of \citet{RTetal06} do not include the PAH emission.
Some Type III YSOs show deep V- or U-shaped SEDs that require
a low dust content within the central $\sim$1,000 AU, indicating
a central hole in the disk.
Using the SED model fits of YSOs and published spectroscopic 
observations of two Type III YSOs, we find at least 11 O-type
YSOs in N\,44.

We have examined the five UCHIIs in N\,44 reported by \citet{IJC04}. 
One is probably a mis-identified background galaxy, and the
remaining four have Type I-II YSO counterparts.  The stellar masses
of these YSOs determined from their SED model fits agree well
with the masses implied by the ionizing fluxes required by
the UCHIIs.  However, the SED model fits suggest envelope 
accretion rates much higher than the critical mass accretion rate
for the formation of UCHIIs.

N\,44 encompasses three molecular concentrations with different
star formation histories.  It is remarkable that the current 
formation of O-type stars is well correlated with the formation 
of such massive stars in the last few Myrs.  The alignment of
YSOs along the southwest rim of N\,44's supershell suggests
that their formation may have been triggered by the expansion of the
supershell.  {\it HST} images show that some YSOs in N\,44 are
in bright-rimmed dust pillars and PDRs with PAH emission, 
indicating that the thermal pressure raised by photo-ionization
or photo-dissociation may have triggered the star formation.

We caution that small \hii\ regions may be associated
with YSOs and thus their dust emission may contaminate the SEDs,
causing uncertainties in the model fits.  {\it HST} images
are needed to resolve such small \hii\ regions. 
Finally, we have analyzed the photometric measurements 
and ionizing power of the central star of N\,44A, and 
we find the star to be O9\,I star, instead of an AGB star
as suggested previously.

\acknowledgments
We thank Leslie Looney for discussion on star formation,
Remy Indebetouw on comparisons with YSOs in the SAGE catalog,
and Barbara Whitney on SED fits.  We also thank Armin Rest's
help on taking CTIO 4m MOSAIC images.  This work was supported
through NASA grants JPL 1264494 ({\it Spitzer Space Telescope}) 
and HST-AR-10942.01-A ({\it Hubble Space Telescope}).
CHRC acknowledges support in part from JPL grants 1282653 and 
1288328 (University of Virginia; PI: Indebetouw).  This study 
made use of data products of the Two Micron All Sky Survey, 
which is a joint project of the University of Massachusetts and 
the Infrared Processing and Analysis Center/California Institute 
of Technology, funded by the National Aeronautics and the Space 
Administration and the National Science Foundation.

%%%%%%%%%%%%%%%%% References %%%%%%%%%%%%%%%%%
%\bibliographystyle{apj}
%\bibliography{refs}

%%%%%%%%%%%%%%%%% Figures %%%%%%%%%%%%%%%%%
\clearpage
\newpage

\begin{figure}
%\plotone{fig1.eps}
\caption[]{MCELS \ha\ and
 MOSAIC B-band images of N\,44.
 (a) MCELS \ha\ image of N\,44, showing the nebular components A-M defined
 by \citet{He56} and the OB associations LH47/48/49 cataloged by \citet{LH70}.
 (b) MOSAIC B-band image of N\,44, showing the OB associations LH47/48/49.
 (c) X-ray contours \citep{CYetal93} overlaid on the MCELS \ha\ image.
 (d) CO contours \citep{FYetal01} overlaid on the MCELS \ha\ image.
 \label{fig:n44opt}}
\end{figure}

\begin{figure}
\epsscale{0.8}
%\plotone{fig2.eps}
\caption[]{IRAC and MIPS images of N\,44.
 (a) 3.6 \um\ image showing stars and modest PAH emission, with
     OB associations LH47/48/49 labeled;
 (b) 8.0 \um\ image showing PAH and dust emissions, (c) 24 \um\ image
 showing dust emission, and (d) color composite of 3.6, 8.0, and 24 \um\
 images.  Dust shrouded objects, e.g., YSOs and AGB stars, appear red. 
 \label{fig:n44img}}
\end{figure}

\begin{figure}
\epsscale{0.8}
%\plotone{fig3.eps}
\caption[]{(a) [8.0] vs.\ [4.5]$-$[8.0]
 CMD of all sources detected in N\,44.  Known AGB stars are marked with
 additional open cyan squares and expected loci from AGB stellar models
 \citep{Gr06} with filled yellow squares.  The criterion to exclude normal
 and AGB stars is shown in short-dashed lines and that to exclude background
 galaxies in long-dashed lines. (b) 99 YSO candidates are found in the upper
 right wedge that has the minimum contamination from stars and background
 galaxies.  These candidates have been through detail examination using
 multi-wavelength images and SEDs.  Candidates that are most likely YSOs are
 marked with additional red open circles and non-YSOs with green crosses.
 \label{fig:cmds}}
\end{figure}

\epsscale{0.4}
\begin{figure}
\plotone{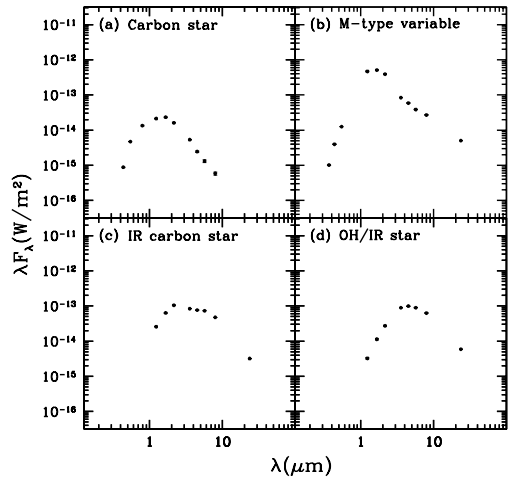}
\caption[]{Selected SEDs of known
 AGB stars in N\,44. (a) Carbon star LMC-BM 24-8 \citep{BM90}, (b) M-type
 variable GRV 0523$-$6752 \citep{RGC88}, (c) IR carbon star MSX LMC 511
 \citep{EVP01}, and (d) OH/IR star MSX LMC 516 \citep{EVP01}.
 \label{fig:agbseds}}
\end{figure} 

\begin{figure}
\epsscale{1.}
%\plotone{fig5.eps}
\caption{(a) YSO candidates in N\,44 from two studies plotted over an
 8 \um\ image.  60 YSO candidates from this study are marked in red
 open circles or question mark, while 19 YSO candidates from 
 \citet{WBetal08} are marked in green open boxes and labeled with
 numbers 1--19.  (b) YSO candidates in N\,44 from this study and
 \citet[][abbreviated as W08 in the figure]{WBetal08} marked on a [8.0]
 vs. ([4.5]-[8.0]) CMD.  The latter are labeled with the same numbers as
 in (a). (c) SEDs of 6 YSO candidates from \citet{WBetal08} but not included
 in the YSO list from this study.  Two numbers are listed for each YSO
 candidate: the ones outside parentheses are serial numbers as labeled in
 (a) and (b), and the ones inside parenthesis are from \citet{WBetal08}. 
 \label{SAGE_comp}}
\end{figure}

\epsscale{0.75}
\begin{figure}
%\plotone{fig6a.eps}
\caption[]{Example YSOs (open circle) in N\,44 shown in multi-wavelength
 images, CMD, and SED to demonstrate our classification scheme.
 (a) Single-source Type I YSO, (b) single-source Type II YSO,
 (c) single-source Type III YSO, (d) YSO in a multiple system, and 
 (e) YSO at the peak of a dust pillar.
 \label{fig:ysoimg}}
\end{figure} 

\clearpage
%{\plotone{fig6b.eps}}\\[5mm]
\centerline{Fig.~\ref{fig:ysoimg} --- Continued.}

\clearpage
%{\plotone{fig6c.eps}}\\[5mm]
\centerline{Fig.~\ref{fig:ysoimg} --- Continued.}

\clearpage
%{\plotone{fig6d.eps}}\\[5mm]
\centerline{Fig.~\ref{fig:ysoimg} --- Continued.}

\clearpage
%{\plotone{fig6e.eps}}\\[5mm]
\centerline{Fig.~\ref{fig:ysoimg} --- Continued.}

\clearpage
\begin{figure}
\epsscale{0.8}
\plotone{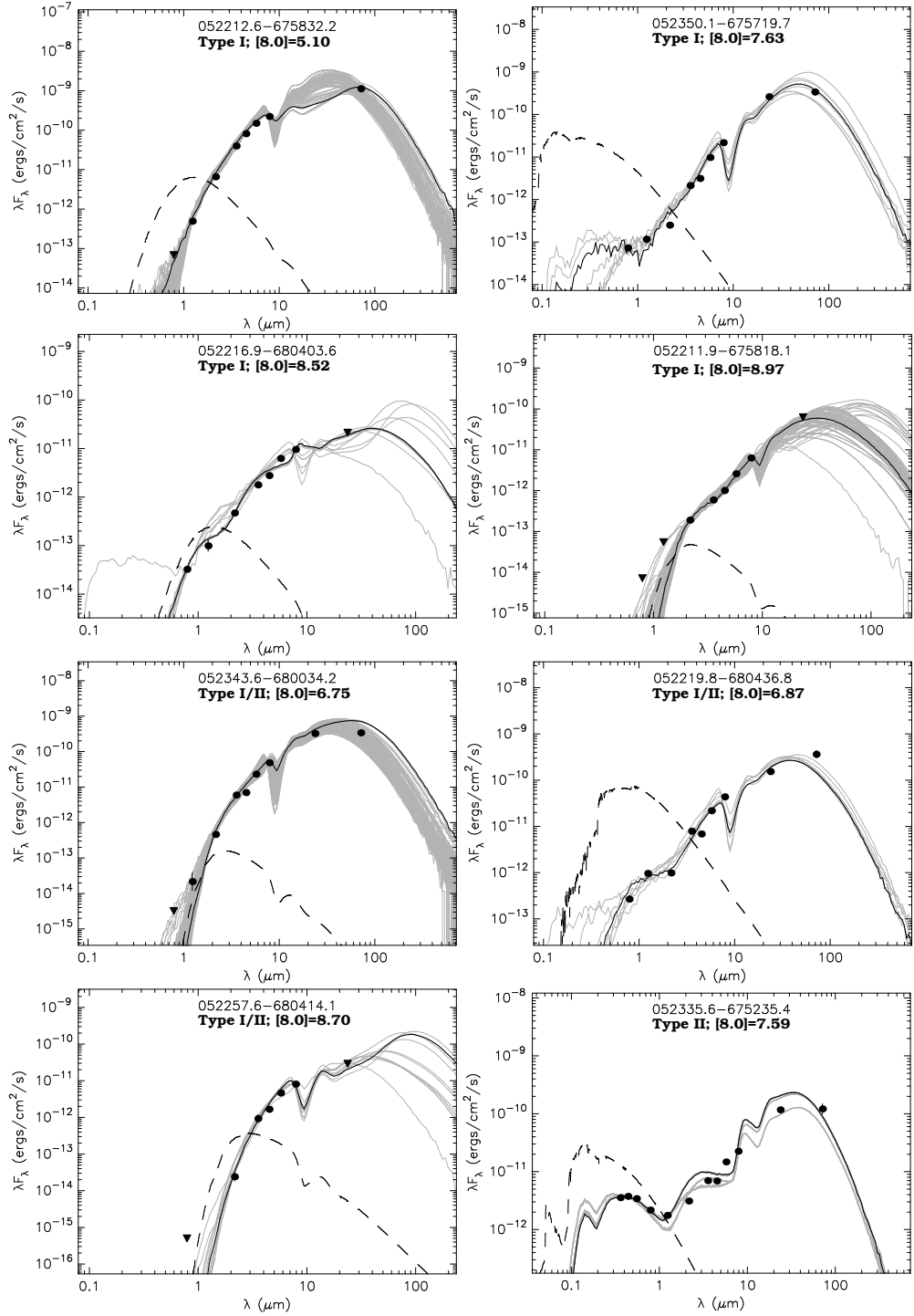}
\caption[SEDs of 36 YSOs analyzed in this study.]{SEDs of 36 YSOs analyzed
 in this study.  
 Filled circles are the flux values converted from magnitudes listed in 
 Table~\ref{ysoclass}.  The source name, type from our empirical 
 classification, and [8.0] mag are labeled at the top of the plot.
 Triangles are upper limits.  Error bars are shown if larger than the data
 points. The solid black line shows the best-fit model, and the dashed black
 line illustrates the radiation from the central star reddened by the best-fit
 $A_V$.  The gray lines  show all acceptable models.  \label{fig:fit}}.
\end{figure}

\clearpage
{\plotone{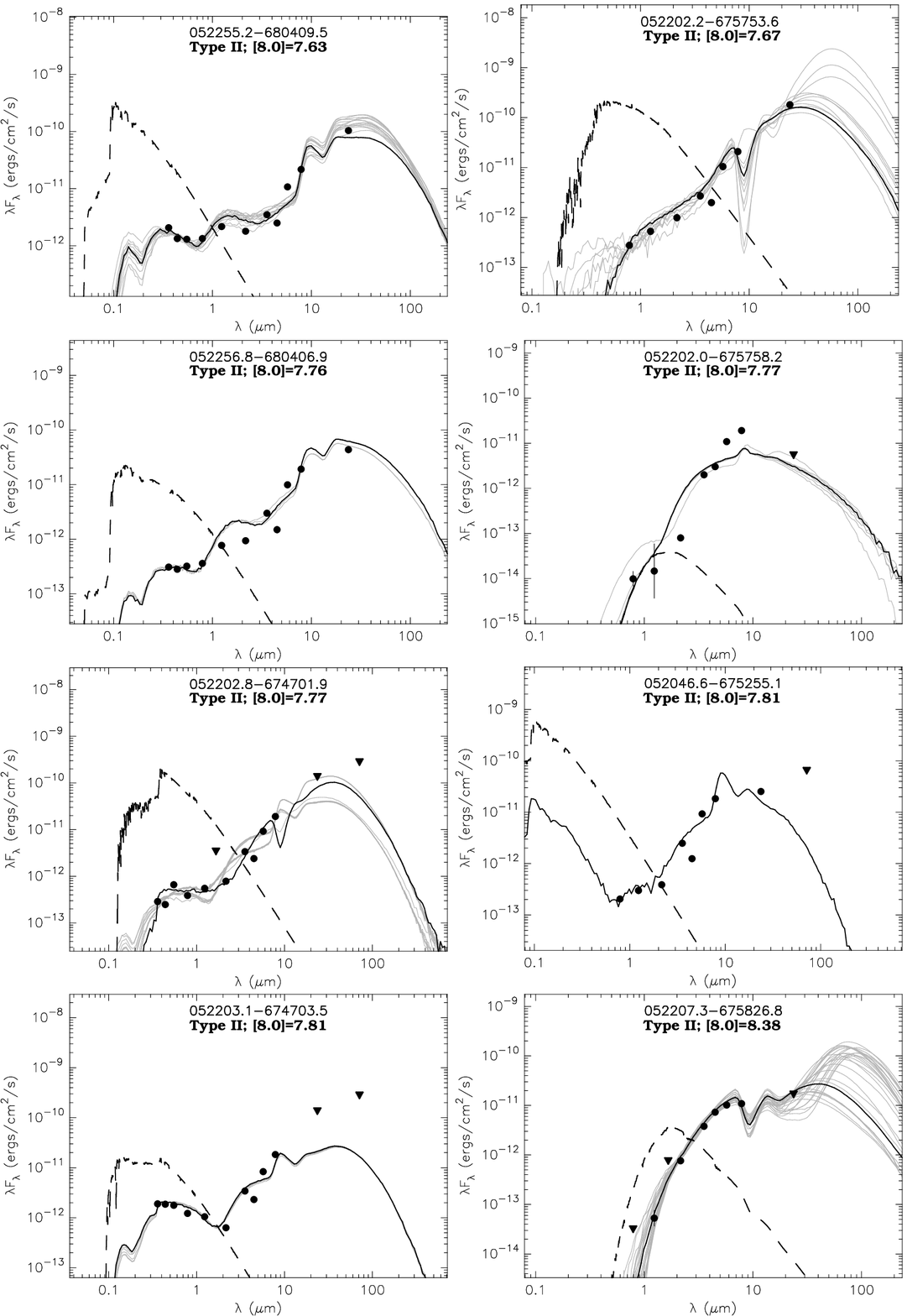}}\\[5mm]
\centerline{Fig.~\ref{fig:fit} --- Continued.}

\clearpage
{\plotone{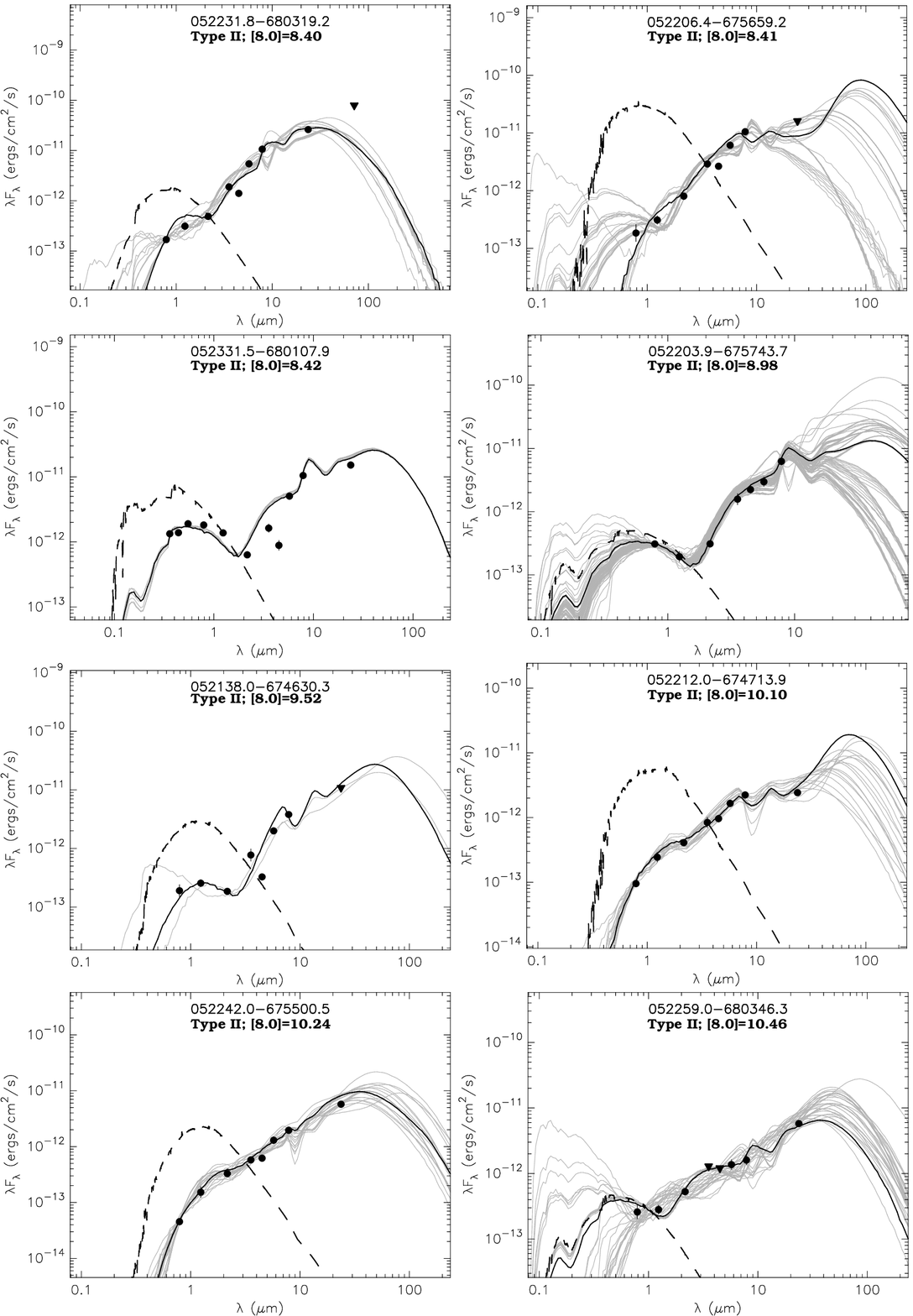}}\\[5mm]
\centerline{Fig.~\ref{fig:fit} --- Continued.}

\clearpage
{\plotone{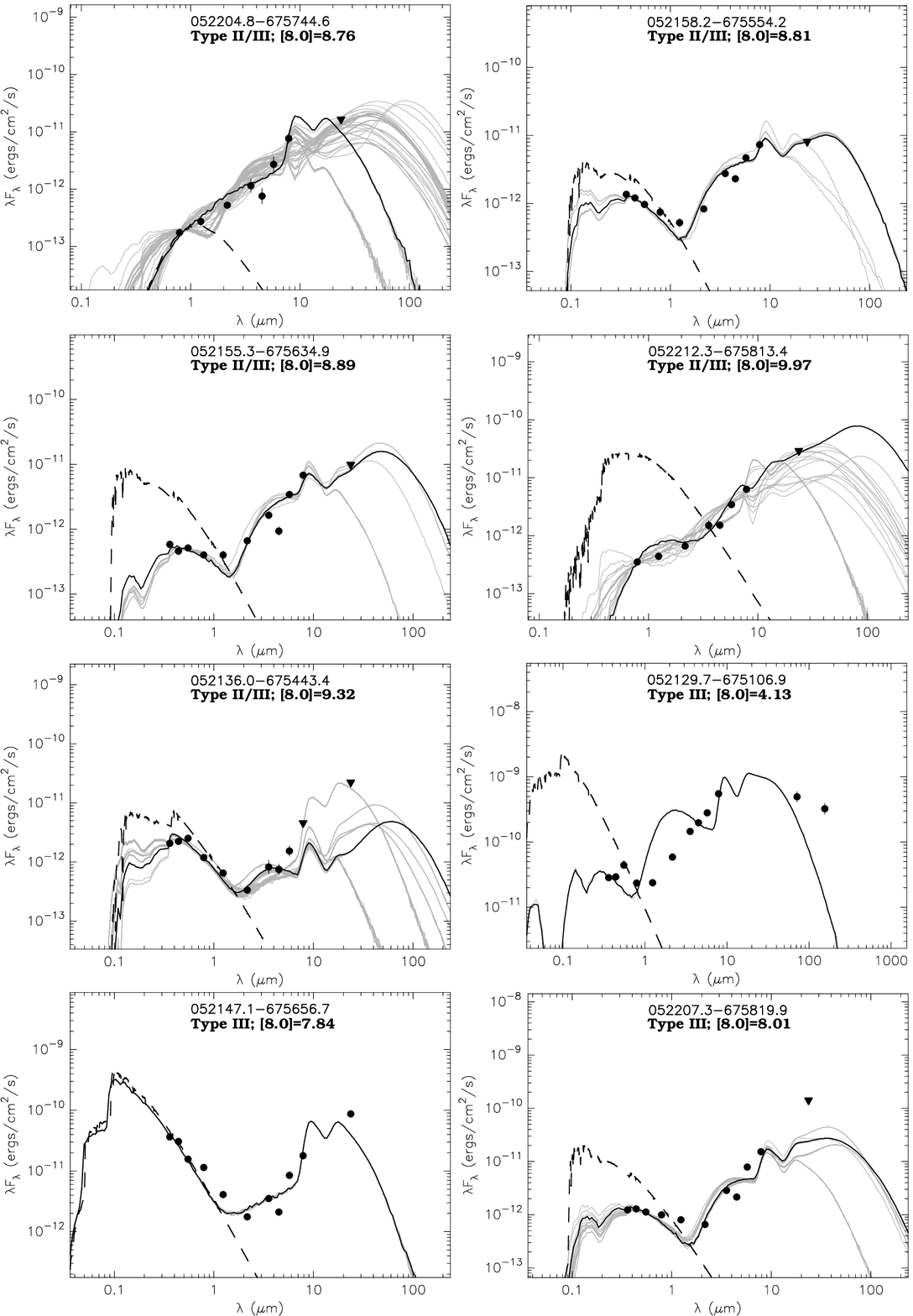}}\\[5mm]
\centerline{Fig.~\ref{fig:fit} --- Continued.}

\clearpage
{\plotone{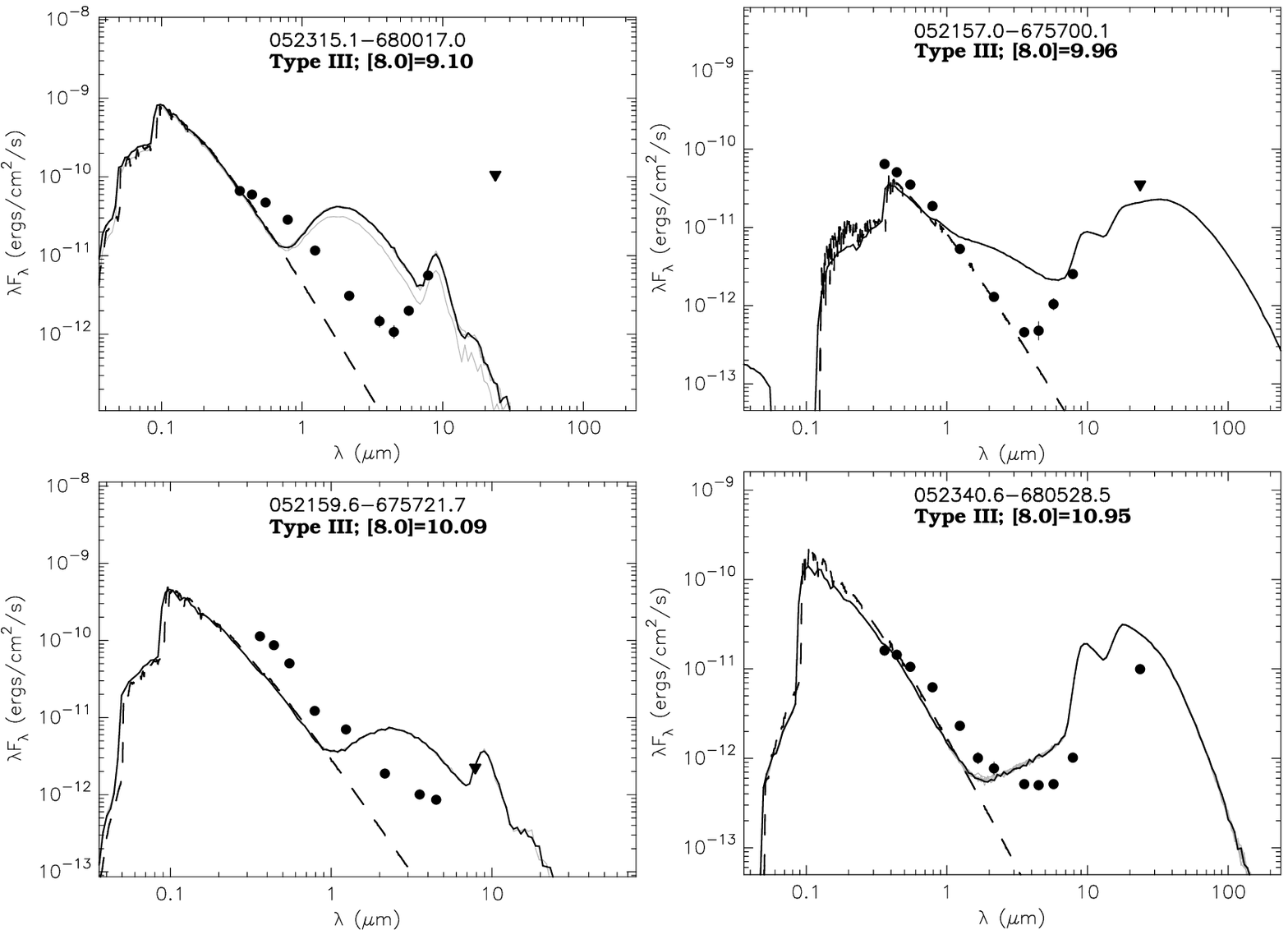}}\\[5mm]
\centerline{Fig.~\ref{fig:fit} --- Continued.}

\begin{figure}
\epsscale{0.8}
\plotone{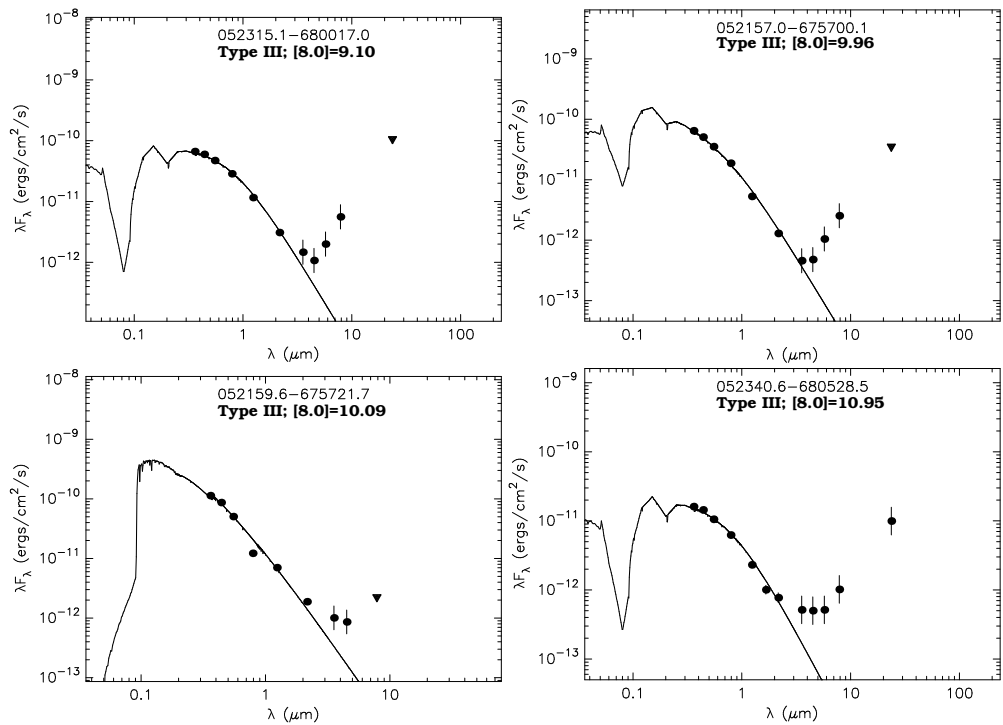}
\caption[]{SEDs of 4 Type III YSOs fitted with stellar atmosphere models 
 \citep{Ku93}. Symbols are the same as Fig.~\ref{fig:fit}. \label{fig:sfit}}.
\end{figure} 

\begin{figure}
\epsscale{.8}
\plotone{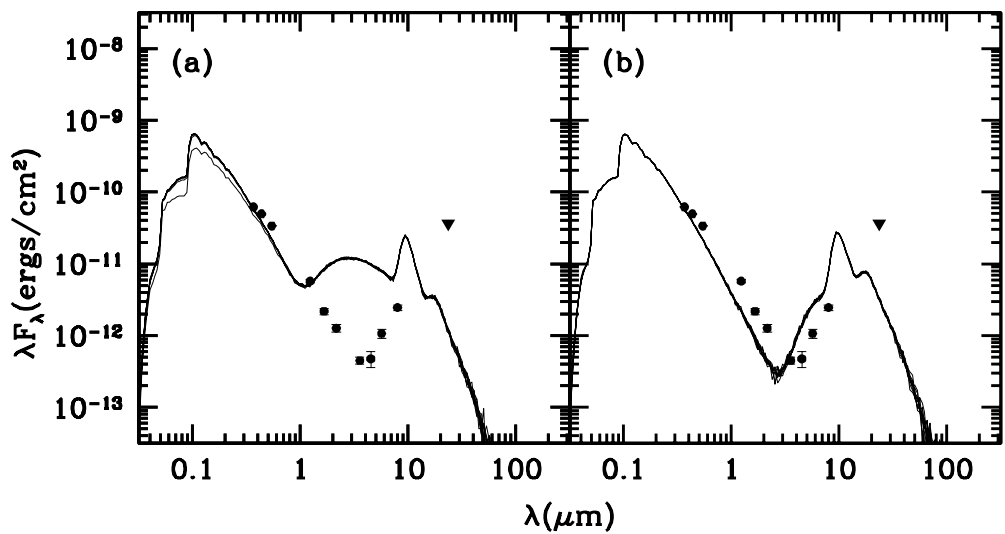}
\caption[]{Model SEDs of Type III YSO 052157.0$-$675700.1 with 
 inner disk radii set as (a) 50 AU and (b) 1000 AU.
 Symbols are the same as Fig.~\ref{fig:fit}. \label{fig:fit_hole}}.
\end{figure}

\epsscale{.9}
\begin{figure}
%\plotone{fig10.eps}
\caption[]{Distribution of YSOs with respect to stellar and
 interstellar environments of N\,44.  The \ha\ image of N\,44 is shown
 in grey scale, overlaid with CO contours (blue lines).
 YSOs with different mass estimates are marked with different symbols:
 O-type, i.e., $M_\star$ Range always $\ge$ 17 $M_\odot$, as large red open 
 circles; early-B type, i.e., $M_\star$ Range always $\ge$ 8 $M_\odot$, as 
 medium red open circles; and B-type, i.e., $M_\star$ Range extending from
 intermediate to high-mass, as small red open circles.
 YSOs without mass estimates are shown as red pluses.  The YSO candidate
 in N\,44A has a yet-to-be-determined nature (see \S7.5) and is shown as
 the green open triangle. \label{fig:yso_pos}}
\end{figure}

\epsscale{0.65}
\begin{figure}
%\plotone{fig11.eps}
\caption[]{YSOs in the \hii\ region N\,44C.
 (a) and (b) show WFPC2 \ha\ and IRAC 8 \um\ images of N\,44C, overlaid
 with positions of YSOs (open circle) and ionizing star (open box). 
 \label{fig:n44c_img}}
\end{figure} 

\begin{figure}
%\plotone{fig12.eps}
\caption[]{YSOs in the \hii\ region N\,44F.
 (a) and (b) show WFPC2 \ha\ and IRAC 8 \um\ images of N\,44C, overlaid
 with positions of YSO (open circle) and ionizing star (open box). 
 \label{fig:n44f_img}}
\end{figure} 

\begin{figure}
%\plotone{fig13.eps}
\caption[YSOs in the \hii\ region N\,44H]{YSOs in the \hii\ region N\,44H.
 (a) and (b) show WFPC2 \ha\ and IRAC 8 \um\ images of N\,44C, overlaid
 with the position of YSO (open circle).
 \label{fig:n44h_img}}
\end{figure} 

\begin{figure}
\plotone{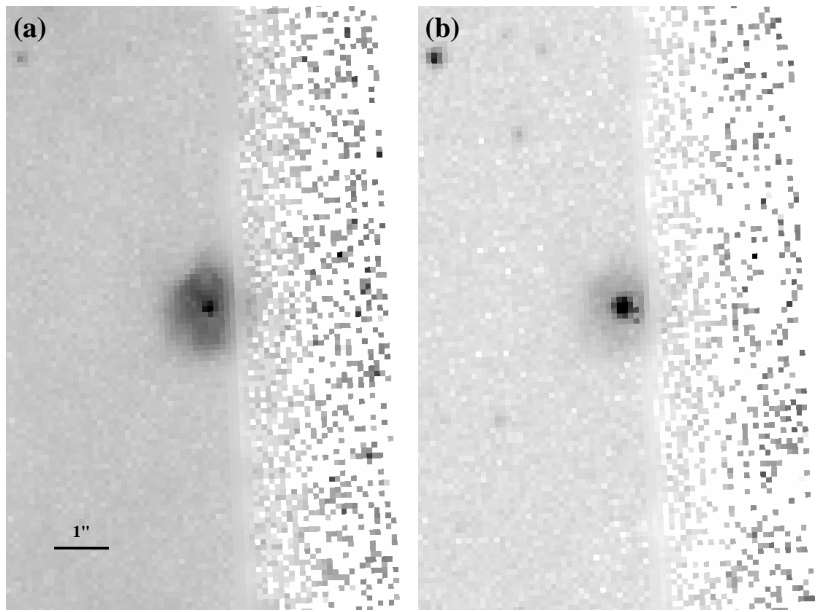}
\caption[]{Close-up
 WFPC2 images of the small \hii\ region associated with YSO
 052207.3$-$675819.9 projected in the outskirt of N\,44C.  The \hii\
 region is shown in (a) \ha\ (F656N) and (b) $y$  (F547M) images.
 \label{fig:n44c_hii}}
\end{figure} 

\epsscale{0.75}
\begin{figure}
%\plotone{fig15.eps}
\caption{YSO 052129.7$-$675106.9 (open circle) in N\,44A shown in
 multi-wavelength images, CMD, and SED.  This bright IR source,
 known as IRAS 05216$-$6753, was previously identified as obscured
 supergiant or AGB star \citep{WPetal92,LCetal97} but the presence
 of compact \hii\ region and the $UBV$ photometry suggest otherwise
 (see text).
 \label{fig:agb2yso}}
\end{figure}

%%%%%%%%%%%%%%%%% Tables %%%%%%%%%%%%%%%%%
\clearpage
\newpage

\begin{deluxetable}{rcccc}
\tablecolumns{5}
\tablecaption{Parameters for IRAC and MIPS Photometric 
Measurements \label{photpar}}
\tablewidth{0pc}
\tablehead{
\colhead{} & \colhead{Aperture} & \colhead{Background} & 
 \multicolumn{1}{c}{Aperture} & \multicolumn{1}{c}{Zero-Mag.} \\
\colhead{} & \colhead{Radius} & \colhead{Annulus} & 
 \multicolumn{1}{c}{Correction} & \multicolumn{1}{c}{Flux} \\
\colhead{Band} & \colhead{($''$)} & \colhead{($''$)} & 
 \multicolumn{1}{c}{Factor} & \multicolumn{1}{c}{(Jy)} 
}
\startdata

IRAC 3.6 \um\ & 3.6 &  3.6-8.4  & 1.124 & 277.5 \\
 4.5 \um\ & 3.6 &  3.6-8.4      & 1.127 & 179.5 \\
 5.8 \um\ & 3.6 &  3.6-8.4      & 1.143 & 116.6 \\
 8.0 \um\ &  3.6 &  3.6-8.4     & 1.234 & ~\,63.1 \\
MIPS~ 24 \um\  & 6~  &  20-32  & 1.699 & ~\,7.14 \\
 70 \um       & 16~~ &  39-65  & 2.087 & 0.775 \\
160 \um       & 40~~ &  75-125 & 1.884 & 0.159 \\

\enddata
\end{deluxetable}

\begin{deluxetable}{ccccccccccccccccr}
\rotate
\tabletypesize{\scriptsize}
\tablecolumns{17}
\tablecaption{Multi-wavelength Photometry of IRAC Sources in N\,44 \label{photcat}}
\tablewidth{0pc}
\tablehead{
 \colhead{Name} &  \colhead{R.A.} & \colhead{Decl.} & 
 \colhead{$U$} & 
 \colhead{$B$} & 
 \colhead{$V$} & 
 \colhead{$I$} & 
 \colhead{$J$} & 
 \colhead{$H$} & 
 \colhead{$K_s$} & 
 \colhead{[3.6]} & 
 \colhead{[4.5]} & 
 \colhead{[5.8]} & 
 \colhead{[8.0]} & 
 \colhead{[24]}  & 
 \colhead{[70]}  &
 \colhead{Flag} \\
}

\startdata

052030.6-680316.7 & 80.12759 & -68.05464 & \nodata \nodata &  19.68   0.03 &  17.86   0.04 &  15.95   0.04 &  14.63   0.04 &  13.82   0.05 &  13.52   0.05 &  13.18   0.09 &  13.39   0.15 &  13.24   0.03 &  12.83   0.06 & \nodata \nodata & \nodata \nodata &  10 \\ 
052030.7-674817.8 & 80.12783 & -67.80495 & \nodata \nodata & \nodata \nodata & \nodata \nodata & \nodata \nodata & \nodata \nodata & \nodata \nodata & \nodata \nodata & \nodata \nodata & \nodata \nodata & \nodata \nodata &  14.46   0.21 & \nodata \nodata & \nodata \nodata &  10 \\ 
052030.7-680353.3 & 80.12783 & -68.06480 & \nodata \nodata &  21.44   0.12 &  20.61   0.12 &  19.30   0.07 & \nodata \nodata & \nodata \nodata & \nodata \nodata &  17.08   0.07 & \nodata \nodata &  16.41   0.47 & \nodata \nodata & \nodata \nodata & \nodata \nodata &  10 \\ 
052030.7-674946.3 & 80.12787 & -67.82953 & \nodata \nodata &  20.57   0.05 &  19.69   0.04 & \nodata \nodata & \nodata \nodata & \nodata \nodata & \nodata \nodata &  16.83   0.06 &  16.78   0.09 & \nodata \nodata & \nodata \nodata & \nodata \nodata & \nodata \nodata &   0 \\ 
052030.7-674436.7 & 80.12789 & -67.74352 &  21.49   0.45 &  20.57   0.06 &  19.46   0.05 &  18.44   0.06 & \nodata \nodata & \nodata \nodata & \nodata \nodata & \nodata \nodata &  17.20   0.10 & \nodata \nodata & \nodata \nodata & \nodata \nodata & \nodata \nodata &  10 \\ 
052030.7-675920.8 & 80.12793 & -67.98911 &  19.46   0.05 &  18.31   0.01 &  17.05   0.02 & \nodata \nodata &  14.89   0.04 &  14.20   0.07 &  14.10   0.07 &  13.90   0.17 &  13.94   0.01 &  13.81   0.04 &  13.50   0.08 & \nodata \nodata & \nodata \nodata &   0 \\ 

\enddata
\tablecomments{Table~\ref{photcat} is presented in its entirety in the
 electronic edition of the Astrophysical Journal.  A portion of is shown
 here for guidance regarding its form and content.  Column 1: source name.
 Columns 2-3: right ascension and declination in degrees.  Columns 4-16:
 photometric measurements of 13 bands from $U$ to 70 \um\ in magnitudes.
 Column 17: data used for $UBVI$ photometry: 0 - $UBV$ from \citet{Ch07},
 10 - $UBVI$ from MCPS.}
\end{deluxetable}

\begin{deluxetable}{rrllc}
\tablecolumns{5}
\tablecaption{Archival {\it HST} WFPC2 Observations of Fields in N\,44
 \label{wfpc2obs}} 
\tablewidth{0pc}
\tablehead{
\colhead{R.A.} & \colhead{Decl.}
& \colhead{} & \colhead{} & 
\multicolumn{1}{c}{Exp. Time}\\
\colhead{(J2000)} & \colhead{(J2000)} & \multicolumn{1}{c}{Program ID/PI} & \multicolumn{1}{c}{Filter\tablenotemark{a}} & 
\colhead{(s)}
}

\startdata

05 21 45.3 & -68 06 13.5 & 7553/MacKenty & F300W & $1\times2900$ \\
05 21 49.8 & -67 54 59.1 & 6698/Chu & F656N & $2\times500$ \\
 & & & F673N & $2\times500$ \\
05 21 58.4 & -68 06 36.2 & 7553/MacKenty & F300W & $1\times2900$ \\
05 22 07.9 & -67 59 13.0 & 6623/Garnett & F502N & $3\times700$ \\
  & & & F547M & $2\times80$ \\
  & & & F656N & $3\times700$ \\
  & & & F675W & $2\times80$ \\
05 22 35.7 & -67 58 12.8 & 6698/Chu & F656N & $2\times500$ \\
 & & & F673N & $2\times500$ \\
05 22 54.9 & -68 01 41.9 & 6540/Schulte-Ladbeck & F656N & $2\times5$, $2\times30$, $2\times500$ \\
05 23 40.2 & -67 59 46.8 & 6253/MacKenty & F300W & $1\times160$, $1\times260$\\

\enddata
\tablenotetext{a}{F330W: Wide $U$; F502N: \oiii; F547M: Str\"omgren $y$; F656N: \ha; F673N: \sii; F675W: WFPC2 $R$.}

\end{deluxetable}

\begin{deluxetable}{cccccccc}
\tablecolumns{8}
\tablecaption{Effective Wavelength and Zero-Magnitude Flux of $UBVIJHK_s$
 Bands \label{flux2mag}} 
\tablewidth{0pc}
\tablehead{
\colhead{} & \colhead{$U$} & \colhead{$B$} & 
\colhead{$V$} & \colhead{$I$} & \colhead{$J$} & \colhead{$H$} & 
\colhead{$K_s$} 
}

\startdata

Effective & & & & & & & \\
Wavelength (\um)& 0.367 & 0.436 & 0.545 & 0.797 & 1.235 & 1.662 & 2.159 \\ 
0$^{th}$ mag & & & & & & & \\
$F_\nu$ (Jy) & 1780 & 4000 & 3600 & 2420 & 1594 & 1024 & 666.8 \\

\enddata
\tablecomments{Adopted from \citet{BM00} for $UBVI$, \citet{CWM03} 
for $JHK_s$.}

\end{deluxetable}

\begin{deluxetable}{rrrrrrrrrll}
\rotate
\tabletypesize{\scriptsize}
\tablecolumns{11}
\tablecaption{Multi-wavelength Photometry of YSO Candidates Selected from CMD
 Criteria \label{ysoclass}}
\tablewidth{0pc}
\tablehead{
 \colhead{Name} &  \colhead{No}  &
 \colhead{$U$} & 
 \colhead{$B$} & 
 \colhead{$V$} & 
 \colhead{$I$} & 
 \colhead{$J$} & 
 \colhead{$H$} & 
 \colhead{$K_s$} & \colhead{} & \colhead{}\\
 \colhead{(1)} &  \colhead{(2)}  &
 \colhead{(3)} & 
 \colhead{(4)} & 
 \colhead{(5)} & 
 \colhead{(6)} & 
 \colhead{(7)} & 
 \colhead{(8)} & 
 \colhead{(9)} & \colhead{} & \colhead{}\\
\colhead{} & \colhead{} &
 \colhead{[3.6]} & 
 \colhead{[4.5]} & 
 \colhead{[5.8]} & 
 \colhead{[8.0]} & 
 \colhead{[24]}  & 
 \colhead{[70]}  &
 \colhead{Flag}  &
 \colhead{Class.}  &
 \colhead{Remarks} \\ 
 \colhead{} &  \colhead{}  &
 \colhead{(10)} & 
 \colhead{(11)} & 
 \colhead{(12)} & 
 \colhead{(13)} & 
 \colhead{(14)} & 
 \colhead{(15)} & 
 \colhead{(16)} & \colhead{(17)} & \colhead{(18)}
}

\startdata

052042.0-674307.7 &	96 & 	   ...   ... &    ...   ... &    ...   ... &    ...   ... &  16.06  0.11 &  15.04  0.11 &  14.71  0.12 && \\ &&  13.85  0.01 &  13.81  0.01 &  13.61  0.04 &  11.64  0.02 &   9.30  0.25 &    ...   ... &  0 & 	G	&		\\
052046.6-675255.1 &	15 & 	   ...   ... &    ...   ... &    ...   ... &  19.23  0.01 &  17.78  0.12 &    ...   ... &  15.95  0.06 && \\ &&  12.45  0.01 &  12.46  0.08 &   9.53  0.02 &   7.81  0.01 &   3.88  0.01 &  -0.79  0.10 & 11 & 	T2	&		\\
052106.8-675715.9 &	99 & 	 19.76  0.17 &  20.29  0.06 &  18.85  0.16 &    ...   ... &  16.61  0.18 &  15.44  0.16 &  15.18  0.20 && \\ &&  14.23  0.02 &  13.99  0.01 &    ...   ... &  11.79  0.03 &    ...   ... &    ...   ... &  0 & 	G  	&		\\
052117.5-680204.6 &	41 & 	   ...   ... &    ...   ... &    ...   ... &  19.66  0.02 &  17.85  0.15 &    ...   ... &  15.98  0.07 && \\ &&  13.36  0.01 &  13.25  0.01 &  10.52  0.04 &   8.73  0.02 &   5.40  0.01 &  -0.26  99.9 & 11 & 	G? YSC?	&		\\
052120.7-674706.6 &	31 & 	   ...   ... &    ...   ... &    ...   ... &    ...   ... &    ...   ... &    ...   ... &    ...   ... && \\ &&  13.03  0.02 &  12.96  0.03 &  10.21  0.02 &   8.51  0.02 &   5.34  0.01 &    ...   ... &  0 & 	D	&		\\
052120.7-674725.5 &	74 & 	 17.39  0.07 &  18.50  0.06 &  17.81  0.08 &  17.25  0.08 &    ...   ... &    ...   ... &    ...   ... && \\ &&  13.65  0.09 &  13.18  0.07 &  11.30  0.12 &   9.54  0.13 &    ...   ... &    ...   ... &  0 & 	S 	&	on IR D 	\\
052120.9-674716.7 &	82 & 	   ...   ... &    ...   ... &    ...   ... &    ...   ... &    ...   ... &    ...   ... &    ...   ... && \\ &&  13.96  0.07 &  13.20  0.04 &    ...   ... &  10.40  0.21 &    ...   ... &    ...   ... &  0 & 	D	&		\\
052120.9-680217.6 &	98 & 	   ...   ... &  22.50  0.50 &  21.91  0.18 &  19.94  0.11 &    ...   ... &    ...   ... &    ...   ... && \\ &&  14.95  0.04 &  13.94  0.02 &  12.70  0.23 &  11.68  0.22 &   7.80  99.9 &    ...   ... &  0 & 	ES?	&		\\
052122.0-674729.0 &	5 & 	   ...   ... &    ...   ... &    ...   ... &  18.77  0.01 &  17.12  0.12 &    ...   ... &  14.88  0.04 && \\ &&  12.08  0.06 &  11.76  0.06 &   9.27  0.03 &   7.44  0.01 &   1.86  0.01 &  -2.17  0.06 & 11 & 	YSO+RN?	&		\\
052126.2-674742.1 &	71 & 	 17.62  0.03 &  18.03  0.02 &  17.91  0.02 &  19.43  0.01 &  17.76  0.15 &    ...   ... &  16.39  0.07 && \\ &&  13.54  0.03 &  13.38  0.02 &    ...   ... &   9.47  0.05 &   5.92  99.9 &    ...   ... & 11 & 	T2, mul	&	in MC	\\
052127.2-675915.1 &	92 & 	 15.37  0.01 &  15.43  0.01 &  14.86  0.01 &    ...   ... &  13.71  0.02 &  13.38  0.04 &  13.35  0.04 && \\ &&  13.08  0.09 &  13.38  0.01 &  12.50  0.07 &  11.24  0.09 &    ...   ... &    ...   ... &  0 & 	blue S	&		\\
052129.7-675106.9 &	1 & 	 14.32  0.01 &  14.96  0.01 &  14.17  0.01 &  14.07  0.01 &  13.00  0.03 &  12.02  0.03 &  10.46  0.02 && \\ &&   8.03  0.01 &   6.96  0.01 &   5.83  0.01 &   4.12  0.01 &   sat.   ... &  -2.95  0.04 & 10 & 	ES	&		\\
052133.3-674420.8 &	97 & 	 16.36  0.07 &  17.11  0.02 &  16.85  0.12 &  17.93  0.01 &  16.52  0.06 &    ...   ... &  15.21  0.05 && \\ &&    ...   ... &  13.82  0.01 &    ...   ... &  11.65  0.05 &   7.98  0.01 &    ...   ... & 11 & 	mul YSOs?	&		\\
052135.5-675500.2 &	86 & 	   ...   ... &  18.49  0.02 &  17.29  0.01 &    ...   ... &  15.09  0.06 &  14.34  0.06 &  14.20  0.08 && \\ &&  14.04  0.26 &  13.94  0.10 &  12.75  0.15 &  10.89  0.13 &    ...   ... &    ...   ... &  0 & 	S	&	in DR	\\
052136.0-675443.4 &	65 & 	 17.16  0.05 &  17.75  0.02 &  17.28  0.04 &  17.31  0.01 &  16.93  0.06 &    ...   ... &  16.10  0.03 && \\ &&  13.65  0.27 &  13.01  0.20 &  11.48  0.19 &   9.32  0.07 &    ...   ... &    ...   ... & 11 & 	T2/3	& tip of DC		\\
052136.3-674643.2 &	89 & 	   ...   ... &    ...   ... &    ...   ... &  21.49  0.12 &  20.48  0.67 &    ...   ... &  17.54  0.08 && \\ &&  15.30  0.14 &  13.65  0.05 &  12.49  0.09 &  11.06  0.16 &   5.36  0.01 &    ...   ... & 11 & 	mul YSOs	&		\\
052136.6-675449.4 &	75 & 	   ...   ... &    ...   ... &    ...   ... &    ...   ... &    ...   ... &    ...   ... &    ...   ... && \\ &&  13.85  0.10 &  13.87  0.17 &  11.44  0.15 &   9.54  0.06 &   4.04  0.01 &    ...   ... &  0 & 	D	&		\\
052138.0-674630.3 &	72 & 	   ...   ... &    ...   ... &    ...   ... &  19.30  0.24 &  17.95  0.11 &    ...   ... &  16.75  0.03 && \\ &&  13.72  0.24 &  13.91  0.07 &  11.20  0.09 &   9.52  0.04 &   4.80  0.01 &    ...   ... & 11 & 	T2	&		\\
052141.9-675324.1 &	63 & 	 20.40  0.25 &  20.42  0.10 &  19.68  0.13 &  19.62  0.02 &  18.13  0.17 &    ...   ... &  16.47  0.05 && \\ &&  13.90  0.05 &  13.81  0.04 &  11.00  0.07 &   9.26  0.04 &   5.32  0.02 &    ...   ... & 11 & 	T2, mul?	&		\\
052144.5-674541.5 &	45 & 	   ...   ... &  21.01  0.10 &  19.88  0.13 &  20.58  0.03 &  19.15  0.20 &    ...   ... &  15.41  0.04 && \\ &&  13.38  0.01 &  13.36  0.01 &  10.65  0.01 &   8.91  0.01 &   5.87  0.03 &   0.37  99.9 & 11 & 	YSC?  	&		\\
052147.1-675656.7 &	17 & 	 14.05  0.01 &  14.91  0.01 &  15.30  0.01 &  14.86  0.02 &  14.94  0.03 &    ...   ... &  14.31  0.03 && \\ &&  12.07  0.04 &  11.89  0.06 &   9.63  0.02 &   7.84  0.01 &   2.54  0.01 &    ...   ... & 11 & 	T3	&		\\
052152.8-675449.5 &	91 & 	 13.98  0.01 &  14.94  0.01 &  14.97  0.01 &    ...   ... &  14.69  0.06 &  14.43  0.07 &  14.25  0.10 && \\ &&  13.68  0.18 &  13.39  0.19 &  12.62  0.08 &  11.18  0.11 &    ...   ... &    ...   ... &  0 & 	S	&		\\
052154.2-674737.1 &	90 & 	   ...   ... &    ...   ... &    ...   ... &  21.54  0.04 &  19.73  0.33 &    ...   ... &  16.58  0.06 && \\ &&  14.41  0.03 &  13.50  0.23 &  12.54  0.21 &  11.10  0.10 &    ...   ... &    ...   ... & 11 & 	T2, mul? 	&	in DC	\\
052155.3-674730.2 &	61 & 	   ...   ... &    ...   ... &    ...   ... &  20.36  0.02 &  18.48  0.18 &    ...   ... &  16.14  0.01 && \\ &&  12.90  0.09 &  11.77  0.04 &  10.49  0.04 &   9.20  0.02 &   5.00  0.01 &    ...   ... & 11 & 	T2, mul	&		\\
052155.3-675634.9 &	46 & 	 18.54  0.07 &  19.47  0.03 &  19.01  0.04 &  18.49  0.01 &  17.47  0.12 &    ...   ... &  15.36  0.01 && \\ &&  12.90  0.10 &  12.77  0.13 &  10.61  0.06 &   8.89  0.04 &   4.90  99.9 &    ...   ... & 11 & 	T2/3	&		\\
052157.0-675700.1 &	77 & 	 13.43  0.01 &  14.36  0.01 &  14.42  0.01 &  14.32  0.01 &  14.66  0.03 &    ...   ... &  14.64  0.01 && \\ &&  14.29  0.12 &  13.50  0.27 &  11.90  0.15 &   9.96  0.10 &   3.52  99.9 &    ...   ... & 11 & 	T3: O8.5V+N	&		\\
052157.5-675618.5 &	59 & 	   ...   ... &    ...   ... &    ...   ... &    ...   ... &    ...   ... &    ...   ... &    ...   ... && \\ &&  13.60  0.10 &  13.40  0.09 &  10.95  0.10 &   9.16  0.07 &   4.27  0.05 &    ...   ... &  0 & 	D	&		\\
052157.9-675625.5 &	68 & 	 11.96  0.01 &  12.95  0.01 &  13.12  0.01 &    ...   ... &  13.32  0.03 &  13.28  0.04 &  13.16  0.04 && \\ &&  12.57  0.10 &  12.45  0.12 &  10.83  0.09 &   9.37  0.08 &    ...   ... &    ...   ... &  0 & 	S: O7III((f))	&		\\
052158.2-675554.2 &	44 & 	 17.62  0.03 &  18.42  0.02 &  18.32  0.02 &  17.82  0.02 &  17.18  0.11 &    ...   ... &  15.12  0.01 && \\ &&  12.34  0.07 &  11.79  0.07 &  10.27  0.05 &   8.81  0.04 &   5.12  99.9 &    ...   ... & 11 & 	T2/3	&		\\
052159.0-674437.2 &	76 & 	 19.57  0.15 &  19.55  0.07 &  18.81  0.10 &  18.55  0.01 &  17.57  0.14 &    ...   ... &  16.36  0.08 && \\ &&  14.13  0.02 &  13.79  0.01 &  11.43  0.01 &   9.71  0.01 &   6.68  0.04 &    ...   ... & 11 & 	T2, mul?	&		\\
052159.6-675721.7 &	79 & 	 12.82  0.01 &  13.78  0.01 &  14.03  0.01 &  14.78  0.02 &  14.35  0.03 &    ...   ... &  14.23  0.03 && \\ &&  13.43  0.10 &  12.86  0.10 &    ...   ... &  10.09  0.24 &    ...   ... &    ...   ... & 11 & 	T3: O7.5V+N	&		\\
052159.6-675715.6 &	78 & 	   ...   ... &    ...   ... &    ...   ... &    ...   ... &    ...   ... &    ...   ... &    ...   ... && \\ &&  14.32  0.36 &  13.90  0.30 &  11.89  0.27 &  10.05  0.26 &    ...   ... &    ...   ... &  0 & 	D	&		\\
052200.4-675745.0 &	25 & 	   ...   ... &    ...   ... &    ...   ... &    ...   ... &    ...   ... &    ...   ... &    ...   ... && \\ &&  13.01  0.13 &  13.49  0.33 &  10.23  0.11 &   8.37  0.11 &    ...   ... &    ...   ... &  0 & 	D	&		\\
052201.9-675732.5 &	56 & 	   ...   ... &  17.37  0.11 &  17.00  0.23 &  19.48  0.12 &  18.05  0.23 &    ...   ... &  16.17  0.10 && \\ &&  13.59  0.33 &  13.39  0.36 &  10.57  0.21 &   9.07  0.25 &    ...   ... &    ...   ... & 11 & 	YSO, mul?	&	in DC	\\
052202.0-675758.2 &	13 & 	   ...   ... &    ...   ... &    ...   ... &  22.52  0.39 &  21.06 99.9 &    ...   ... &  17.66  0.15 && \\ &&  12.69  0.12 &  11.50  0.08 &   9.36  0.05 &   7.77  0.03 &    ...   ... &    ...   ... & 11 & 	T2	&	in DC	\\
052202.2-675753.6 &	11 & 	   ...   ... &    ...   ... &    ...   ... &  18.89  0.02 &  17.16  0.11 &    ...   ... &  14.93  0.04 && \\ &&  12.36  0.11 &  11.95  0.12 &   9.40  0.04 &   7.67  0.03 &   1.75  0.02 &    ...   ... & 11 & 	T2	&		\\
052202.3-674657.5 &	33 & 	   ...   ... &    ...   ... &    ...   ... &  20.81  0.10 &  18.28  0.19 &    ...   ... &  15.96  0.04 && \\ &&  13.18  0.17 &  12.86  0.03 &  10.35  0.06 &   8.51  0.03 &   2.41  0.01 &    ...   ... & 11 & 	YSO 	&	in DC	\\
052202.8-674701.9 &	14 & 	 19.31  0.14 &  20.14  0.13 &  18.74  0.14 &  18.53  0.01 &  17.11 99.9 &    ...   ... &  15.18  0.01 && \\ &&  12.12  0.06 &  11.75  0.07 &   9.54  0.03 &   7.77  0.02 &   2.01  99.9 &  -2.39  99.9 & 11 & 	T2	&	at tip of DC	\\
052203.1-674703.5 &	16 & 	 17.26  0.10 &  17.95  0.05 &  17.66  0.11 &  17.28  0.01 &  16.42  0.07 &    ...   ... &  15.41  0.03 && \\ &&  12.10  0.02 &  11.78  0.02 &   9.64  0.03 &   7.81  0.03 &    ...   ... &    ...   ... & 11 & 	T2	&		\\
052203.4-675746.9 &	24 & 	   ...   ... &    ...   ... &    ...   ... &    ...   ... &    ...   ... &    ...   ... &  16.58  0.09 && \\ &&  12.76  0.14 &  11.78  0.08 &  10.01  0.07 &   8.28  0.04 &   1.74  99.9 &    ...   ... & 11 & 	T1/2?	&	in DC	\\
052203.9-675743.7 &	52 & 	   ...   ... &    ...   ... &    ...   ... &  18.78  0.02 &  18.25  0.19 &    ...   ... &  16.19  0.07 && \\ &&  12.95  0.20 &  11.82  0.10 &  10.77  0.17 &   8.98  0.10 &    ...   ... &    ...   ... & 11 & 	T2 	&	in DC  	\\
052204.1-674709.7 &	73 & 	 17.74  0.06 &  18.39  0.05 &  18.48  0.06 &  18.14  0.12 &    ...   ... &    ...   ... &    ...   ... && \\ &&  13.92  0.04 &  13.94  0.04 &  11.24  0.04 &   9.52  0.05 &    ...   ... &    ...   ... &  0 & 	S 	&	on IR D 	\\
052204.8-675744.6 &	42 & 	   ...   ... &    ...   ... &    ...   ... &  19.40  0.03 &  17.88  0.14 &    ...   ... &  15.62  0.05 && \\ &&  13.29  0.27 &  13.00  0.31 &  10.87  0.33 &   8.76  0.20 &    ...   ... &    ...   ... & 11 & 	T2/3 	&	at edge of DC	\\
052204.9-675720.6 &	62 & 	   ...   ... &    ...   ... &    ...   ... &    ...   ... &    ...   ... &    ...   ... &    ...   ... && \\ &&  13.45  0.12 &  13.42  0.12 &  11.16  0.23 &   9.22  0.20 &    ...   ... &    ...   ... &  0 & 	D	&		\\
052204.9-675801.6 &	36 & 	 20.70  0.23 &  19.87  0.05 &  19.00  0.03 &    ...   ... &    ...   ... &    ...   ... &    ...   ... && \\ &&  13.08  0.16 &  12.61  0.11 &  10.25  0.14 &   8.54  0.06 &   2.08  0.02 &    ...   ... &  0 & 	D	&		\\
052205.2-675741.6 &	38 & 	   ...   ... &    ...   ... &    ...   ... &  22.18  0.37 &  20.08  0.64 &    ...   ... &  16.25  0.08 && \\ &&  13.39  0.25 &  12.79  0.15 &    ...   ... &   8.61  0.23 &    ...   ... &    ...   ... & 11 & 	mul YSOs &	in DC	\\
052205.3-675748.5 &	21 & 	   ...   ... &    ...   ... &    ...   ... &    ...   ... &    ...   ... &    ...   ... &  17.22  0.13 && \\ &&  13.12  0.26 &  12.97  0.35 &   9.88  0.06 &   8.13  0.04 &   2.06  99.9 &    ...   ... & 11 & 	T2?	&	in DC 	\\
052206.4-675659.2 &	28 & 	   ...   ... &    ...   ... &    ...   ... &  19.34  0.36 &  17.74  0.15 &    ...   ... &  15.16  0.04 && \\ &&  12.28  0.07 &  11.64  0.06 &   9.98  0.04 &   8.41  0.02 &   4.37  99.9 &    ...   ... & 11 & 	T2	&		\\
052207.3-675819.9 &	19 & 	 17.73  0.01 &  18.35  0.01 &  18.16  0.01 &  17.51  0.01 &  16.71  0.09 &    ...   ... &  15.38  0.05 && \\ &&  12.30  0.07 &  11.87  0.07 &   9.71  0.03 &   8.01  0.02 &   2.79  99.9 &    ...   ... & 11 & 	T3	&	in HII	\\
052207.3-675826.8 &	26 & 	   ...   ... &    ...   ... &    ...   ... &  21.20 99.9 &  19.66 99.9 &    ...   ... &  15.21  0.04 && \\ &&  11.99  0.07 &  10.54  0.02 &   9.43  0.03 &   8.38  0.04 &   3.05  99.9 &    ...   ... & 11 & 	T2	&	in DC	\\
052207.7-675649.4 &	85 & 	 18.68  0.08 &  18.91  0.04 &  19.03  0.04 &    ...   ... &    ...   ... &    ...   ... &    ...   ... && \\ &&  14.19  0.07 &  13.94  0.10 &  12.50  0.07 &  10.69  0.07 &    ...   ... &    ...   ... &  0 & 	S	&	in DR	\\
052208.5-675821.3 &	30 & 	   ...   ... &    ...   ... &    ...   ... &    ...   ... &    ...   ... &    ...   ... &    ...   ... && \\ &&  12.93  0.04 &  12.46  0.04 &  10.16  0.06 &   8.45  0.03 &   2.90  0.04 &    ...   ... &  0 & 	D	&		\\
052208.6-675805.5 &	51 & 	   ...   ... &    ...   ... &    ...   ... &  19.35  0.05 &  17.83  0.15 &    ...   ... &  15.24  0.04 && \\ &&  12.32  0.08 &  11.68  0.08 &  10.42  0.10 &   9.01  0.09 &   4.44  0.25 &    ...   ... & 11 & 	mul YSOs	&		\\
052208.6-675921.9 &	70 & 	 18.00  0.02 &  18.75  0.01 &  18.68  0.02 &  18.57  0.01 &  17.61  0.14 &    ...   ... &  16.43  0.09 && \\ &&  13.79  0.04 &  13.83  0.03 &  11.10  0.08 &   9.38  0.05 &   5.92  0.14 &    ...   ... & 11 & 	T2, mul?	&		\\
052208.8-675325.2 &	60 & 	 18.61  0.06 &  18.98  0.03 &  18.68  0.04 &  18.05  0.01 &  17.37  0.11 &    ...   ... &  16.23  0.04 && \\ &&  13.35  0.11 &  13.13  0.15 &  10.79  0.04 &   9.19  0.02 &   5.63  0.02 &    ...   ... & 11 & 	T2 + T3/S	&		\\
052208.9-674703.4 &	93 & 	 17.39  0.08 &  16.74  0.02 &  15.82  0.04 &  14.81  0.04 &  14.11  0.03 &  13.61  0.04 &  13.52  0.04 && \\ &&  13.33  0.01 &  13.38  0.18 &  12.64  0.09 &  11.37  0.14 &    ...   ... &    ...   ... &  0 & 	blue S	&		\\
052211.9-675818.1 &	48 & 	   ...   ... &    ...   ... &    ...   ... &  22.79  0.27 &  19.55  0.31 &    ...   ... &  16.72  0.09 && \\ &&  14.01  0.13 &  12.69  0.17 &  10.92  0.10 &   8.97  0.05 &   2.04  99.9 &    ...   ... & 11 & 	T1 	&	in DC	\\
052212.0-674713.9 &	80 & 	 21.47  0.53 &  21.01  0.11 &  20.55  0.13 &  20.05  0.02 &  18.00  0.15 &    ...   ... &  15.89  0.03 && \\ &&  13.63  0.17 &  12.74  0.10 &  11.40  0.09 &  10.10  0.05 &   6.43  0.01 &    ...   ... & 11 & 	T2	&		\\
052212.3-675813.4 &	49 & 	   ...   ... &    ...   ... &    ...   ... &  18.63  0.01 &  17.35  0.11 &    ...   ... &  15.37  0.04 && \\ &&  12.99  0.17 &  12.24  0.12 &  10.61  0.10 &   8.97  0.07 &    ...   ... &    ...   ... & 11 & 	T2/3 	&	in DC	\\
052212.6-675832.2 &	2 & 	   ...   ... &    ...   ... &    ...   ... &  20.34 99.9 &  17.24  0.12 &    ...   ... &  12.87  0.01 && \\ &&   9.44  0.01 &   7.92  0.01 &   6.51  0.01 &   5.10  0.01 &   sat.    ... &  -3.85  0.08 & 11 & 	T1 	&		\\
052216.7-675837.7 &	55 & 	   ...   ... &    ...   ... &    ...   ... &    ...   ... &    ...   ... &    ...   ... &    ...   ... && \\ &&  13.92  0.14 &  13.57  0.15 &  10.85  0.10 &   9.07  0.09 &    ...   ... &    ...   ... &  0 & 	D	&		\\
052216.8-680428.3 &	66 & 	   ...   ... &    ...   ... &    ...   ... &    ...   ... &    ...   ... &    ...   ... &    ...   ... && \\ &&  13.91  0.06 &  13.92  0.08 &  11.12  0.04 &   9.32  0.05 &   4.96  0.04 &    ...   ... &  0 & 	D peak	&		\\
052216.9-680403.6 &	34 & 	   ...   ... &    ...   ... &    ...   ... &  21.22  0.08 &  18.98  0.27 &    ...   ... &  15.73  0.06 && \\ &&  12.81  0.09 &  11.59  0.04 &   9.96  0.03 &   8.52  0.03 &   4.01  0.01 &    ...   ... & 11 & 	T1	&		\\
052217.8-680432.9 &	69 & 	   ...   ... &    ...   ... &    ...   ... &    ...   ... &    ...   ... &    ...   ... &    ...   ... && \\ &&  13.80  0.06 &  13.68  0.08 &  11.07  0.06 &   9.36  0.06 &    ...   ... &    ...   ... &  0 & 	D	&		\\
052218.9-675813.8 &	64 & 	   ...   ... &    ...   ... &    ...   ... &    ...   ... &    ...   ... &    ...   ... &    ...   ... && \\ &&    ...   ... &  13.84  0.12 &  11.21  0.15 &   9.31  0.13 &    ...   ... &    ...   ... &  0 & 	D	&		\\
052219.8-680436.8 &	4 & 	   ...   ... &  18.24  0.08 &  18.16  0.13 &  18.93  0.14 &  16.51  0.09 &    ...   ... &  14.93  0.04 && \\ &&  11.20  0.04 &  10.61  0.03 &   8.60  0.02 &   6.87  0.01 &   1.93  0.01 &  -2.62  0.07 & 11 & 	T1/2 	&	in DC	\\
052221.0-680515.3 &	94 & 	   ...   ... &    ...   ... &    ...   ... &    ...   ... &    ...   ... &    ...   ... &    ...   ... && \\ &&  15.35  0.02 &  13.85  0.28 &  12.81  0.26 &  11.38  0.11 &   8.32  0.21 &    ...   ... &  0 & 	ES?	&		\\
052227.7-675412.8 &	37 & 	   ...   ... &    ...   ... &    ...   ... &  21.85  0.22 &  18.85  0.24 &    ...   ... &  15.92  0.04 && \\ &&  13.05  0.13 &  12.95  0.15 &  10.32  0.06 &   8.55  0.03 &   4.20  0.04 &    ...   ... & 11 & 	T2? 	&	in DC	\\
052229.1-675339.5 &	50 & 	 16.11  0.03 &  17.04  0.04 &  16.65  0.02 &  16.21  0.01 &  15.95  0.06 &    ...   ... &  15.12  0.02 && \\ &&  13.08  0.10 &  13.05  0.15 &  10.68  0.05 &   9.00  0.02 &   5.34  0.03 &    ...   ... & 11 & 	T3, mul 	&		\\
052230.4-675443.9 &	18 & 	 16.11  0.01 &  16.75  0.01 &  16.40  0.01 &    ...   ... &  14.84  0.08 &  14.42  0.10 &  13.62  0.11 && \\ &&  11.93  0.06 &  11.38  0.06 &    ...   ... &   7.88  0.09 &   2.38  0.01 &  -2.69  0.09 &  0 & 	S 	&	in HII	\\
052231.8-680319.2 &	27 & 	 18.75  0.10 &  19.65  0.06 &  19.59  0.06 &  19.44  0.03 &  17.73  0.14 &    ...   ... &  15.70  0.06 && \\ &&  12.75  0.10 &  12.33  0.09 &  10.11  0.05 &   8.40  0.03 &   3.85  0.01 &    ...   ... & 11 & 	T2	&	in DC?  	\\
052232.7-680301.7 &	53 & 	   ...   ... &    ...   ... &    ...   ... &  19.45  0.01 &  17.26  0.11 &    ...   ... &  15.06  0.04 && \\ &&  13.06  0.10 &  12.69  0.09 &  10.90  0.07 &   9.02  0.03 &   4.42  0.01 &    ...   ... & 11 & 	T2, mul 	&	in DC  	\\
052241.4-675508.2 &	84 & 	 16.15  0.01 &  16.19  0.01 &  15.55  0.01 &    ...   ... &  14.12  0.03 &  13.62  0.04 &  13.57  0.05 && \\ &&  13.37  0.12 &  13.33  0.17 &  12.02  0.16 &  10.52  0.06 &    ...   ... &    ...   ... &  0 & 	S	&		\\
052242.0-675500.5 &	81 & 	   ...   ... &    ...   ... &    ...   ... &  20.86  0.03 &  18.51  0.22 &    ...   ... &  16.12  0.01 && \\ &&  14.04  0.04 &  13.21  0.02 &  11.66  0.13 &  10.24  0.10 &   5.50  0.02 &    ...   ... & 11 & 	T2 	&		\\
052249.0-680111.1 &	67 & 	   ...   ... &    ...   ... &    ...   ... &    ...   ... &    ...   ... &    ...   ... &    ...   ... && \\ &&  13.93  0.16 &  13.76  0.11 &  11.20  0.16 &   9.36  0.09 &    ...   ... &    ...   ... &  0 & 	D	&		\\
052249.2-680129.0 &	23 & 	 17.66  0.02 &  18.41  0.02 &  17.98  0.02 &  15.84  0.03 &  15.88  0.02 &    ...   ... &  14.77  0.03 && \\ &&  12.51  0.07 &  12.34  0.09 &   9.97  0.01 &   8.26  0.02 &   4.27  0.01 &    ...   ... & 11 & 	T2, mul	&		\\
052251.0-680401.0 &	35 & 	   ...   ... &    ...   ... &    ...   ... &    ...   ... &    ...   ... &    ...   ... &    ...   ... && \\ &&  13.16  0.08 &  13.21  0.06 &  10.32  0.07 &   8.52  0.04 &    ...   ... &    ...   ... &  0 & 	D	&		\\
052251.7-680436.7 &	39 & 	 18.35  0.06 &  19.07  0.02 &  18.76  0.03 &    ...   ... &    ...   ... &    ...   ... &    ...   ... && \\ &&  12.98  0.11 &  12.80  0.11 &  10.41  0.09 &   8.63  0.06 &    ...   ... &    ...   ... &  0 & 	S	&	near DC	\\
052253.7-680434.8 &	20 & 	   ...   ... &    ...   ... &    ...   ... &    ...   ... &    ...   ... &    ...   ... &    ...   ... && \\ &&  12.85  0.14 &  12.57  0.16 &   9.82  0.06 &   8.12  0.04 &   2.01  0.02 &    ...   ... &  0 & 	D	&		\\
052254.6-680424.3 &	10 & 	   ...   ... &    ...   ... &    ...   ... &    ...   ... &    ...   ... &    ...   ... &    ...   ... && \\ &&  12.04  0.06 &  11.90  0.08 &   9.52  0.10 &   7.65  0.03 &    ...   ... &    ...   ... &  0 & 	D 	&		\\
052254.9-680422.2 &	6 & 	   ...   ... &    ...   ... &    ...   ... &    ...   ... &    ...   ... &    ...   ... &    ...   ... && \\ &&  11.93  0.05 &  11.67  0.07 &   9.32  0.04 &   7.49  0.07 &   1.64  0.01 &    ...   ... &  0 & 	D	&		\\
052255.2-680409.5 &	8 & 	 17.16  0.06 &  18.31  0.04 &  18.00  0.04 &  17.18  0.01 &  15.63  0.05 &    ...   ... &  14.28  0.03 && \\ &&  12.08  0.07 &  11.70  0.07 &   9.37  0.03 &   7.63  0.02 &   2.35  0.01 &    ...   ... & 11 & 	T2	&		\\
052255.4-680431.6 &	22 & 	   ...   ... &  20.69  0.03 &  20.15  0.03 &    ...   ... &    ...   ... &    ...   ... &    ...   ... && \\ &&  13.15  0.12 &  13.05  0.14 &    ...   ... &   8.25  0.12 &    ...   ... &    ...   ... &  0 & 	D	&		\\
052256.8-680406.9 &	12 & 	 19.23  0.07 &  20.00  0.04 &  19.52  0.05 &  18.61  0.01 &  16.75  0.09 &    ...   ... &  14.99  0.04 && \\ &&  12.25  0.09 &  12.26  0.12 &   9.46  0.04 &   7.76  0.02 &   3.29  0.01 &    ...   ... & 11 & 	T2	&		\\
052257.6-680414.1 &	40 & 	   ...   ... &    ...   ... &    ...   ... &    ...   ... &    ...   ... &    ...   ... &  18.97  0.28 && \\ &&  13.52  0.24 &  12.14  0.09 &  10.28  0.06 &   8.70  0.04 &   3.65  99.9 &    ...   ... & 11 & 	T1/2  	&		\\
052259.0-680346.3 &	83 & 	   ...   ... &    ...   ... &    ...   ... &  18.97  0.24 &  17.84  0.14 &    ...   ... &  15.61  0.05 && \\ &&  13.16  99.9 &  12.49  99.9 &  11.61  0.16 &  10.46  0.14 &   5.49  0.05 &    ...   ... & 11 & 	T2 	&		\\
052308.7-680006.8 &	43 & 	   ...   ... &    ...   ... &    ...   ... &  20.50  0.03 &  18.69  0.18 &    ...   ... &  16.28  0.04 && \\ &&  13.53  0.17 &  12.94  0.15 &  10.61  0.07 &   8.78  0.04 &   3.51  0.01 &    ...   ... & 11 & 	YSC	&	in DC  	\\
052309.4-680205.5 &	58 & 	 18.48  0.01 &  18.91  0.01 &  18.63  0.01 &    ...   ... &    ...   ... &    ...   ... &    ...   ... && \\ &&  13.69  0.04 &  13.78  0.04 &  10.88  0.05 &   9.15  0.05 &   6.04  0.02 &    ...   ... &  0 & 	D peak	&		\\
052311.4-680040.9 &	54 & 	   ...   ... &    ...   ... &    ...   ... &    ...   ... &    ...   ... &    ...   ... &    ...   ... && \\ &&  13.52  0.06 &  13.41  0.04 &  10.81  0.06 &   9.07  0.05 &   4.62  0.03 &    ...   ... &  0 & 	D peak  	&		\\
052315.1-680017.0 &	57 & 	 13.40  0.01 &  14.19  0.01 &  14.10  0.01 &  13.86  0.01 &  13.81  0.04 &    ...   ... &  13.69  0.05 && \\ &&  13.02  0.17 &  12.62  0.17 &  11.20  0.12 &   9.10  0.06 &   2.32  0.01 &    ...   ... & 11 & 	T3	&	in HII	\\
052318.0-675938.6 &	32 & 	   ...   ... &    ...   ... &    ...   ... &    ...   ... &  16.34  0.14 &  15.94  0.20 &  14.70 99.9 && \\ &&  13.15  0.10 &  12.73  0.07 &  10.25  0.06 &   8.51  0.04 &    ...   ... &    ...   ... &  0 & 	D	&		\\
052318.0-675942.8 &	47 & 	   ...   ... &    ...   ... &    ...   ... &    ...   ... &    ...   ... &    ...   ... &    ...   ... && \\ &&  13.49  0.16 &  12.94  0.24 &  10.59  0.09 &   8.94  0.06 &    ...   ... &    ...   ... &  0 & 	D peak	&		\\
052331.5-680107.9 &	29 & 	 17.64  0.03 &  18.27  0.02 &  17.59  0.02 &  16.85  0.01 &  16.12  0.06 &    ...   ... &  15.41  0.05 && \\ &&  12.91  0.12 &  12.83  0.16 &  10.19  0.06 &   8.42  0.04 &   4.44  0.01 &    ...   ... & 11 & 	T2	&		\\
052335.6-675235.4 &	7 & 	 16.58  0.02 &  17.19  0.03 &  16.95  0.02 &  16.66  0.01 &  15.86  0.06 &    ...   ... &  13.69  0.02 && \\ &&  11.33  0.02 &  10.60  0.02 &   9.03  0.01 &   7.59  0.01 &   2.22  0.01 &  -1.43  0.08 & 11 & 	T2	&		\\
052340.6-680528.5 &	87 & 	 14.94  0.03 &  15.73  0.07 &  15.73  0.02 &  15.52  0.01 &  15.56  0.07 &    ...   ... &  15.20  0.18 && \\ &&  14.16  0.04 &  13.45  0.04 &  12.67  0.05 &  10.95  0.04 &   4.90  0.01 &   0.58  0.23 & 11 & 	T3 or B[e] S	&		\\
052343.6-680034.2 &	3 & 	   ...   ... &    ...   ... &    ...   ... &  23.65  99.9 &  20.63  0.62 &    ...   ... &  15.76  0.05 && \\ &&  11.50  0.04 &  10.59  0.03 &   8.54  0.01 &   6.75  0.01 &   1.12  0.01 &  -2.55  0.07 & 11 & 	T1/2	&		\\
052343.9-680056.1 &	88 & 	   ...   ... &    ...   ... &    ...   ... &  21.66  0.08 &  18.69  0.24 &    ...   ... &  16.97  0.11 && \\ &&  14.16  0.05 &  13.71  0.03 &  12.40  0.09 &  11.00  0.17 &    ...   ... &    ...   ... & 11 & 	mul YSOs	&		\\
052350.1-675719.7 &	9 & 	   ...   ... &    ...   ... &    ...   ... &  20.35  0.04 &  18.81  0.21 &    ...   ... &  16.43  0.07 && \\ &&  12.62  0.09 &  11.46  0.04 &   9.48  0.03 &   7.63  0.01 &   1.35  0.01 &  -2.54  0.06 & 11 & 	T1	&		\\
052351.1-675326.6 &	95 & 	   ...   ... &    ...   ... &    ...   ... &    ...   ... &    ...   ... &    ...   ... &    ...   ... && \\ &&  14.54  0.01 &  13.60  0.22 &  12.52  0.02 &  11.54  0.12 &   9.07  0.07 &    ...   ... &  0 & 	ES?	&		\\

\enddata

\tablecomments{Column 1: source name. Column 2: Ranking of the brightness at
 8 \um. Columns 3-16: photometric measurements of 13 bands from $U$ to 70
 \um\ in magnitudes.  Measurements with uncertainties of $99.9$ are the
 upper brightness limits as they include fluxes
 from neighbors or backgrounds.  Column 17: data used for 
 $IJHK$ photometry: 0 -- $I$ from MCPS and $JHK$ from 2MASS
 catalog, 10 -- $I$ from 4m Mosaic data and $JHK$ from 2MASS
 catalog, 11 -- $I$ from 4m Mosaic data and $JHK$ from 4m ISPI
 data.  Column 18 and 19: classification and remarks: D -- diffuse emission,
 DC -- dark cloud, DR -- dusty region,  ES -- evolved star, G -- background
 galaxy, MC -- molecular cloud, mul -- multiple, N -- nebula, RN -- reflection
 nebula, S -- star, T1/2/3 -- Type I/II/III YSO, YSC -- young star cluster.}

\end{deluxetable}

\begin{deluxetable}{rcrrrll}
\tablecolumns{7}
\tablecaption{YSO Candidates in N\,44 in Two Catalogs \label{wc_id}}
\tablewidth{0pc}
\tablehead{
 \colhead{} & \colhead{W08} &  \colhead{Source Name} & \multicolumn{2}{c}{Offset} & \multicolumn{2}{c}{Classification} \\
 \colhead{No.} & \colhead{No.} &  \colhead{in This Study} & \colhead{$\Delta \alpha$} & \colhead{$\Delta \delta$} & \colhead{W08} & \colhead{This Study} 
}

\startdata

1 & 530 & 052042.1-675255.0 & $-$0.41 &  0.21 & YSO & 	G cut \\ 
2 & 546 & 052116.2-674511.7 & $-$0.09 & $-$0.28 & YSO & 	E cut \\ 
3 & 552 & 052129.7-675106.9 & $-$0.07 &  0.27 & PN & 	III? E? \\ 
4 & 557 & 052147.1-675656.7 & $-$0.34 &  0.07 & YSO\_hp & 	III \\ 
5 & 563 & 052155.3-674730.2 & $-$0.52 & $-$0.13 & YSO\_hp & 	II \\ 
6 & 565 & 052159.0-674437.2 & $-$0.12 &  0.12 & YSO & 	II \\ 
7 & 566 & 052202.0-675758.2 & $-$0.25 &  0.19 & YSO & 	II \\ 
8 & 572 & 052207.3-675826.8 &  0.05 & $-$0.03 & YSO & 	II \\ 
9 & 574 & 052211.5-675401.9 & $-$0.23 &  0.35 & YSO & 	G cut \\ 
10 & 576 & 052212.0-674713.9 & $-$0.44 &  0.01 & YSO\_hp & 	II \\ 
11 & 579 & 052212.6-675832.2 & $-$0.10 & $-$0.06 & YSO & 	I \\ 
12 & 581 & 052216.9-680403.6 & $-$0.28 & $-$0.22 & YSO\_hp & 	I \\ 
13 & 598 & 052249.2-680129.0 & $-$0.61 &  0.10 & YSO & 	II \\ 
14 & 604 & 052259.0-680346.3 & $-$0.37 &  0.23 & YSO & 	II \\ 
15 & 606 & 052302.2-680400.1 & $-$0.47 &  0.34 & YSO & 	G cut \\ 
16 & 609 & 052308.7-680006.8 & $-$0.67 &  0.23 & YSO & 	YSC \\ 
17 & 613 & 052318.5-680045.5 & $-$0.52 & $-$0.19 & YSO & 	E cut \\ 
18 & 616 & 052335.6-675235.4 & $-$0.19 & $-$0.27 & YSO\_hp & 	II \\ 
19 & 622 & 052351.3-680712.3 & $-$0.81 &  0.10 & YSO\_hp & 	E cut \\ 

\enddata
\end{deluxetable}
\begin{deluxetable}{lrcccrrrrrrrrrrl}
\rotate
\tabletypesize{\scriptsize}
\tablecolumns{16}
\tablecaption{Inferred Physical Parameters from SED Fits to YSOs
 \label{sedfits}}
\tablewidth{0pc}
\tablehead{
 \colhead{} &
 \colhead{} &
 \colhead{} &
 \colhead{M$_{\ast}$} &
 \colhead{} &
\multicolumn{10}{c}{Physical Parameters of the Best-Fit Model} & \colhead{}\\
 \cline{6-15}
 \colhead{} & \colhead{} &  \colhead{} & \colhead{} &  \colhead{} &
 \colhead{} & \colhead{} &  \colhead{} & \colhead{} &  \colhead{} &
 \colhead{} & \colhead{} \\
 \colhead{} & 
 \colhead{[8.0]} & 
 \colhead{} & 
 \colhead{Range} & 
 \colhead{Stage} & 
 \colhead{M$_{\ast}$} & 
 \colhead{T$_{\ast}$} &
 \colhead{R$_{\ast}$} & 
 \colhead{Age} & 
 \colhead{$\dot{M}_{\rm env}$} & 
 \colhead{$\dot{M}_{\rm disk}$} & 
 \colhead{$M_{\rm disk}$} & 
 \colhead{$i$} & 
 \colhead{$L_{\rm tot}$} &
 \colhead{A$_V$} &
 \colhead{} \\
 \colhead{Source Name} & 
 \colhead{(mag)} & 
 \colhead{Type} & 
 \colhead{($M_\odot$)} & 
 \colhead{Range} & 
 \colhead{($M_\odot$)} & 
 \colhead{(K)} &
 \colhead{($R_\odot$)} & 
 \colhead{(yr)} & 
 \colhead{($M_\odot$/yr)} & 
 \colhead{($M_\odot$/yr)} & 
 \colhead{($M_\odot$)} & 
 \colhead{($\degr$)} & 
 \colhead{($L_\odot$)} &
 \colhead{(mag)} &
 \colhead{Remark} 
}

\startdata

052212.6-675832.2 & 	 5.10 & I       & 	20-50 & 	I & 	26 & 16000 & 37.7 & 8E+03 & 7.6E$-$03 & 0.0E+00 & 0.0E+00 &  18 & 92000 & 6.1 	& DC, UCHII\tablenotemark{a} 	\\
052350.1-675719.7 & 	 7.63 & I       & 	17-25 & 	I & 	21 & 33000 & 7.5 & 3E+04 & 1.2E$-$03 & 0.0E+00 & 0.0E+00 &  56 & 62000 & 1.1 	& DC 	\\
052216.9-680403.6 & 	 8.52 & I       & 	8-15 & 	I,II & 	8 & 16000 & 8.8 & 2E+05 & 1.0E$-$05 & 1.8E$-$07 & 6.0E$-$02 &  41 &  4300 & 7.2 	&  	\\
052211.9-675818.1 & 	 8.97 & I       & 	8-28 & 	I,II & 	28 & 4200 & 300.0 & 1E+03 & 5.9E$-$04 & 0.0E+00 & 0.0E+00 &  56 & 26000 & 2.1 	& DC \\
\cline{1-16}
052343.6-680034.2 & 	 6.75 & I/II    & 	17-37 & 	I,II & 	20 & 18000 & 23.3 & 2E+04 & 5.9E$-$03 & 0.0E+00 & 0.0E+00 &  18 & 53000 & 27.5 	& DC, UCHII\tablenotemark{a} 	\\
052219.8-680436.8 & 	 6.87 & I/II    & 	17-24 & 	I & 	21 & 8000 & 93.7 & 5E+03 & 3.9E$-$04 & 0.0E+00 & 0.0E+00 &  49 & 33000 & 1.9 	& DC, UCHII\tablenotemark{a} 	\\
052257.6-680414.1 & 	 8.70 & I/II    & 	10-16 & 	I,II & 	16 & 8400 & 56.6 & 1E+04 & 6.5E$-$03 & 0.0E+00 & 0.0E+00 &  18 & 15000 & 29.7 	& CO peak 	\\
\cline{1-16}
052335.6-675235.4 & 	 7.59 & II      & 	14-17 & 	II,III & 	17 & 33000 & 5.3 & 9E+05 & 1.7E$-$08 & 6.7E$-$11 & 1.9E$-$06 &  18 & 30000 & 0.9 	&  	\\
052255.2-680409.5 & 	 7.63 & II      & 	13-18 & 	I,II & 	13 & 30000 & 4.6 & 3E+05 & 4.0E$-$06 & 7.8E$-$07 & 3.6E$-$01 &  81 & 15000 & 0.0 	& UCHII\tablenotemark{a}	\\
052202.2-675753.6 & 	 7.67 & II      & 	17-36 & 	I,II & 	18 & 6400 & 125.0 & 6E+03 & 1.4E$-$04 & 8.8E$-$06 & 9.4E$-$02 &  87 & 23000 & 0.2 	& DC? 	\\
052256.8-680406.9 & 	 7.76 & II      & 	11-12 & 	I & 	12 & 29000 & 4.3 & 2E+05 & 1.7E$-$05 & 1.8E$-$08 & 2.3E$-$02 &  87 & 11000 & 0.7 	&  	\\
052202.0-675758.2 & 	 7.77 & II      & 	9 & 	II & 	9 & 24000 & 3.7 & 2E+06 & 0.0E+00 & 6.1E$-$07 & 6.8E$-$02 &  56 & 4100 & 9.9 	& DC 	\\
052202.8-674701.9 & 	 7.77 & II      & 	8-15 & 	I,II,III & 	15 & 8900 & 47.3 & 1E+04 & 2.0E$-$04 & 3.0E$-$06 & 6.9E$-$02 &  49 & 13000 & 0.1 	& DC tip 	\\
052046.6-675255.1 & 	 7.81 & II      & 	17 & 	II & 	17 & 33000 & 5.3 & 1E+06 & 0.0E+00 & 8.7E$-$07 & 1.5E$-$01 &  87 & 32000 & 0.0 	&  	\\
052203.1-674703.5 & 	 7.81 & II      & 	8-14 & 	I,III & 	8 & 16000 & 8.8 & 2E+05 & 1.0E$-$05 & 1.8E$-$07 & 6.0E$-$02 &  41 & 4300 & 0.4 	&  	\\
052207.3-675826.8 & 	 8.38 & II      & 	7-25 & 	I,II & 	11 & 4400 & 76.8 & 5E+03 & 1.8E$-$04 & 1.8E$-$04 & 4.0E$-$01 &  18 & 2700 & 7.8 	& DC 	\\
052231.8-680319.2 & 	 8.40 & II      & 	8-15 & 	I,II & 	8 & 12000 & 13.9 & 1E+05 & 8.9E$-$05 & 5.2E$-$10 & 8.0E$-$04 &  18 & 3200 & 2.9 	& DC? 	\\
052206.4-675659.2 & 	 8.41 & II      & 	8-15 & 	I,II & 	13 & 5900 & 78.2 & 1E+04 & 4.2E$-$03 & 4.0E$-$06 & 2.2E$-$01 &  18 & 6600 & 1.2 	&  	\\
052331.5-680107.9 & 	 8.42 & II      & 	8 & 	I & 	8 & 16000 & 8.8 & 2E+05 & 1.0E$-$05 & 1.8E$-$07 & 6.0E$-$02 &  69 & 4300 & 0.9 	&  	\\
052203.9-675743.7 & 	 8.98 & II      & 	8-15 & 	I,II & 	8 & 22000 & 3.4 & 8E+05 & 5.0E$-$07 & 5.5E$-$06 & 1.2E$-$01 &  31 & 2700 & 2.0 	& DC 	\\
052138.0-674630.3 & 	 9.52 & II      & 	9-10 & 	I & 	9 & 7000 & 31.4 & 4E+04 & 4.5E$-$04 & 1.2E$-$05 & 2.4E$-$02 &  18 & 2200 & 3.5 	&  	\\
052212.0-674713.9 & 	 10.10 & II     & 	2-9 & 	I & 	9 & 5100 & 48.7 & 3E+04 & 1.3E$-$03 & 5.1E$-$07 & 8.7E$-$02 &  18 & 1400 & 1.9 	&  	\\
052242.0-675500.5 & 	 10.24 & II     & 	5-10 & 	I & 	9 & 6200 & 31.4 & 5E+04 & 6.3E$-$05 & 4.8E$-$07 & 6.1E$-$02 &  41 & 1300 & 3.6 	&  	\\
052259.0-680346.3 & 	 10.46 & II     & 	4-11 & 	I,II & 	6 & 18000 & 3.6 & 5E+05 & 1.0E$-$05 & 2.9E$-$08 & 1.7E$-$03 &  18 & 1200 & 1.9 	&  	\\
\cline{1-16}
052204.8-675744.6 & 	 8.76 & II/III  & 	8-33 & 	I,II,III & 	9 & 6000 & 43.2 & 3E+04 & 6.1E$-$05 & 3.9E$-$09 & 2.4E$-$03 &  31 & 2200 & 0.7 	& DC edge 	\\
052158.2-675554.2 & 	 8.81 & II/III  & 	5-14 & 	I,III & 12 & 7100 & 51.8 & 2E+04 & 3.8E$-$03 & 1.7E$-$07 & 2.5E$-$03 &  18 & 6100 & 1.0 	&  	\\
052155.3-675634.9 & 	 8.89 & II/III  & 	8-9 & 	I,II & 	8 & 22000 & 3.4 & 4E+05 & 6.2E$-$05 & 8.2E$-$08 & 1.6E$-$02 &  63 &  2300 & 0.5 	&  	\\
052212.3-675813.4 & 	 8.97 & II/III  & 	5-28 & 	I,II,III & 	12 & 7100 & 51.8 & 2E+04 & 3.8E$-$03 & 1.7E$-$07 & 2.5E$-$03 &  18 & 6100 & 1.0 	& superposed on DC	\\
052136.0-675443.4 & 	 9.32 & II/III  & 	6-11 & 	II,III & 	6 & 13000 & 5.8 & 5E+05 & 2.3E$-$07 & 1.9E$-$09 & 3.2E$-$03 &  41 & 890 & 0.1 	& dust pillar tip 	\\
\cline{1-16}
052129.7-675106.9 & 	 4.12 & III     & 	34 & 	II & 	34 & 42000 & 7.8 & 5E+05 & 1.2E$-$06 & 1.1E$-$10 & 4.6E$-$04 &  41 & 160000 & 0.0 	& $\sim$O9I, this study \\
052147.1-675656.7 & 	 7.84 & III     & 	17 & 	III & 	17 & 33000 & 5.3 & 1E+06 & 0.0E+00 & 8.7E$-$13 & 1.0E$-$06 &  81 & 32000 & 0.1 	&  	\\
052207.3-675819.9 & 	 8.01 & III     & 	9-15 & 	II & 	9 & 25000 & 3.7 & 3E+05 & 3.0E$-$06 & 1.8E$-$08 & 1.1E$-$03 &  41 & 4600 & 0.4 	&  	\\
052315.1-680017.0 & 	 9.10 & III     & 	20-21 & 	III & 	21 & 36000 & 5.9 & 1E+06 & 0.0E+00 & 4.7E$-$12 & 2.4E$-$06 &  75 & 50000 & 0.0 	&  B0-O5V\tablenotemark{b}	\\
052157.0-675700.1 & 	 9.96 & III     & 	9 & 	I & 	9 & 9200 & 20.1 & 6E+04 & 9.6E$-$05 & 9.9E$-$06 & 5.5E$-$01 &  18 & 2800 & 0.0 	& O8.5V\tablenotemark{c}	\\
052159.6-675721.7 & 	 10.09 & III     & 	16 & 	III & 	16 & 32000 & 5.1 & 1E+06 & 0.0E+00 & 1.5E$-$11 & 4.7E$-$07 &  31 & 25000 & 0.0 	& O7.5V\tablenotemark{c}	\\
052340.6-680528.5 & 	 10.95 & III     & 	12 & 	III & 	12 & 28000 & 4.4 & 2E+06 & 0.0E+00 & 9.3E$-$14 & 7.4E$-$07 &  87 & 11000 & 0.0 	& B[e]\tablenotemark{c}	\\

\enddata
\tablenotetext{a}{\citet{IJC04}}
\tablenotetext{b}{\citet{Ch07}}
\tablenotetext{c}{\citet{OM95}}

\end{deluxetable}

\begin{deluxetable}{llcccllcc}
\rotate
\tabletypesize{\scriptsize}
\tablecolumns{9}
\tablecaption{Properties of YSOs with UCHIIs \label{uch2}}
\tablewidth{0pc}
\tablehead{
 \colhead{} & 
 \colhead{} & 
 \colhead{} & 
 \colhead{Stage} & 
 \colhead{$M_\star$} & 
 \colhead{Spec.} &
 \colhead{Spec.\tablenotemark{a}} & 
 \colhead{$\dot{M}_{\rm env}$} &
 \colhead{$\dot{M}_{\rm crit}$\tablenotemark{b}} \\
 \colhead{UCHII} & 
 \colhead{YSO ID} & 
 \colhead{Type} & 
 \colhead{Range} & 
 \colhead{($M_\odot$)} & 
 \colhead{Type} & 
 \colhead{Type} &
 \colhead{($M_\odot$/yr)} & 
 \colhead{($M_\odot$/yr)} 
 }

\startdata

B0522$-$5800     & 052212.6$-$675832.2 & I    & I     & 26 & O7-8 V
 & O7 V   & 7.6E$-$03 & 4.1E$-$05 \\
B0523$-$6806(NE) & 052343.6$-$680034.2 & I/II & I,II  & 20 & O8-9 V
 & O9 V   & 5.9E$-$03 & 1.8E$-$05 \\
B0523$-$6806(SW) & 052219.8$-$680436.8 & I/II & I     & 21 & O8-9 V
 & O8.5 V & 3.9E$-$04 & 2.2E$-$05 \\
B0523$-$6806     & 052255.2$-$680409.5 & II   & I,II  & 13 & B1-2 V
 & B0 V   & 4.0E$-$06 & 1.0E$-$05 \\

\enddata
\tablenotetext{a}{The spectral type is determined from radio observations 
 \citep{IJC04}.}
\tablenotetext{b}{The critical infalling rate is adopted from \citet{CE02}.}

\end{deluxetable}

\end{document}